\documentclass[nofootinbib,twocolumn,pra,aps]{revtex4}

\usepackage[T1]{fontenc} 
\usepackage{latexsym} 
\usepackage{amsmath} 

\usepackage{mathtools}

\usepackage{epsf}
\usepackage{url}
\usepackage{epic,eepic}

\usepackage{amssymb} 
\usepackage{exscale} 
\usepackage{graphicx} 
\usepackage[polish]{} 
\usepackage{makeidx}
\usepackage{enumerate}
\usepackage{color}

\newtheorem{cor}{Corollary}
\newtheorem{remark}{Remark}

\newenvironment{prooftw}[1][Proof]

{\begin{quotation}\begin{tabbing}
            \hspace{1em}\=\hspace{1em}\=\hspace{1em}\=\hspace{1em}\=%
            \hspace{1em}\=\hspace{1em}\=\hspace{1em}\kill}%
        {\end{tabbing}\end{quotation}}

\newtheorem{definition}{Definition}
\newtheorem{theorem}{Theorem}
\newtheorem{lemma}{Lemma}

\newtheorem{proposition}{Proposition}
\newtheorem{observation}{Observation}
\def\qed{$\square$}

\font\Bbb =msbm10

\def\eea{\end{array}}
\def\bea{\begin{array}}

\newcommand{\be}{\begin{equation}}
\newcommand{\ee}{\end{equation}}
\newcommand{\emx}{\end{array}}
\newcommand{\bex}{\begin{eqnarray}}
\newcommand{\enx}{\end{eqnarray}}
\newcommand{\ben}{\begin{enumerate}}
\newcommand{\enn}{\end{enumerate}}
\newcommand{\bei}{\begin{itemize}}
\newcommand{\eei}{\end{itemize}}
\newcommand{\enxn}{\nonumber\end{eqnarray}}
\font\Bbb =msbm10  scaled \magstephalf
\def\id{{\hbox{\Bbb I}}}
\def\1{{\hbox{\Bbb I}}}

\def\duzomniejsze{<\kern-.7mm<}
\def\duzowieksze{>\kern-.7mm>}

\def\beq{\begin{equation}}
\def\eeq{\end{equation}}
\def\be{\begin{equation}}
\def\ee{\end{equation}}
\def\ben{\begin{eqnarray}}
\def\een{\end{eqnarray}}
\def\beqa{\begin{eqnarray}}
\def\eeqa{\end{eqnarray}}
\def\eea{\end{array}}
\def\bea{\begin{array}}

\def\tr{{\rm Tr}}
\def\id{{\rm I}}

\def\>{\rangle}
\def\<{\langle}

\def\ot{\otimes}

\def\dt#1{{{\kern -.0mm\rm d}}#1\,}

\def\CAB{{C\leftrightarrow AB}}

\def\ABC{{A \leftrightarrow C\leftrightarrow B}}

\def\r{\rho}

\def\id{{\mathbb{I}}}

\begin{document}
\title{ Limitations for private randomness repeaters }
\author{Karol Horodecki$^{1,2}$, Ryszard P. Kostecki$^{1}$, Roberto Salazar$^{1}$, and Micha\l{} Studzi\'nski$^{3}$}
\affiliation{$^1$ National Quantum Information Centre in Gda{\'n}sk and Institute of Informatics,
Faculty of Mathematics, Physics and Informatics, University of Gda{\'n}sk, 80-952 Gda{\'n}sk, Poland\\
$^2$ International Centre for Theory of Quantum Technologies,
University of Gda{\'n}sk, 80-952 Gda{\'n}sk, Poland\\
$^3$ National Quantum Information Centre in Gda{\'n}sk and Institute of Theoretical Physics and Astrophysics,
Faculty of Mathematics, Physics and Informatics, University of Gda{\'n}sk, 80-952 Gda{\'n}sk, Poland
}

\begin{abstract}
Cryptographic protocols are often based on the two main resources: private randomness and private key. In this paper, we develop the relationship between these two resources. First, we show that any state containing perfect, directly accessible, private key (a private state) is a particular case of the state containing perfect, directly accessible, private randomness (an independent state). We then demonstrate a fundamental limitation on the possibility of transferring the privacy of random bits in quantum networks with an intermediate repeater station. More precisely, we provide an upper bound on the rate of repeated randomness in this scenario, similar to the one derived for private key repeaters. This bound holds for states with positive partial transposition. We further demonstrate the power of this upper bound by showing a gap between the localisable and the repeated private randomness for separable Werner states. In case of restricted class of operations, we provide also a bound on repeated randomness which holds for arbitrary states.
\end{abstract}

\maketitle
\section{introduction}
Ensuring the security of communication in quantum internet is one of the main current challenges of quantum technology \cite{Kozlowski-Wehner}. In this context, two distant honest parties must distribute a secure key, i.e., a private correlated string of bits.  A prominent security framework that assures the distribution of encrypted bits in a quantum network is the  {\it quantum repeaters} scheme \cite{ent_swapping,BBPSSW1996, repeatersPRL,repeaters}.  It allows for distributing secure key employing pure maximally entangled states \cite{RMPK-quant-ent} and {\it entanglement swapping} \cite{ent_swapping,BBPSSW1996}.

In a recent article \cite{BCHW} the paradigm of network key swapping was extended to the most general scenario of {\it private states} \cite{pptkey,keyhuge}, that are, generally, mixed quantum states. A striking result of \cite{BCHW} is the existence of mixed states $\rho$ and $\rho'$, such that no protocol between three parties $A,B,C=C_1C_2$ can transfer a non-negligible amount of key between $A$ and $B$ from the key shared between the parties $AC_1$ and $BC_2$. 

This fact shows an intriguing property of the secure key extracted from mixed quantum states: it is not {\it transitive} for an arbitrary state, i.e., the fact that $A$ has secure connection with $C$ and $C$ has secure connection with $B$ does not imply that $A$ can establish secure connection with $B$. 

In this article, we investigate the network properties of another critical resource for cryptography: {\it the private randomness}. In most cases, it is used for testing a quantum device or post-processing the classical outcome of the latter. For this reason, the privacy of randomness appears as a precondition for secure key distribution. This resource was recognised quite early (for the review on this topic see \cite{Randomness-review}; the framework for single-party private randomness extraction was developed in \cite{BFW-decoupling}), and has motivated commercial implementations (e.g., \cite{idq}). Only recently a  {\it resource theory} framework of (distributed) private randomness has been established \cite{YHW} (see, e.g., \cite{RMPK-quant-ent,Chitambar-Gour} for a review of other resource theories). According to this approach, the task of distillation of private randomness amounts to obtaining the so-called {\it independent states}
$\alpha$ via {\it closed local operations and dephasing channel} $\mathrm{CLODCC}$.
More precisely the CLODCC operations are compositions of
(i) local unitary operations by each of the honest parties ($U_A$ and $U_B$)
(ii) communication via {\it dephasing channel} from $A$ to $B$ and vice versa.
The dephasing channel transfers the state measured in a fixed (say computational) basis. These operations were introduced in context of purity distillation.
The choice of this class of operations in resource theory of private randomness is justified, as these operations do not bring in private randomness.
\begin{figure}
    \centering
    \includegraphics[trim= 1cm 7cm 0cm 0cm,width=1\columnwidth]{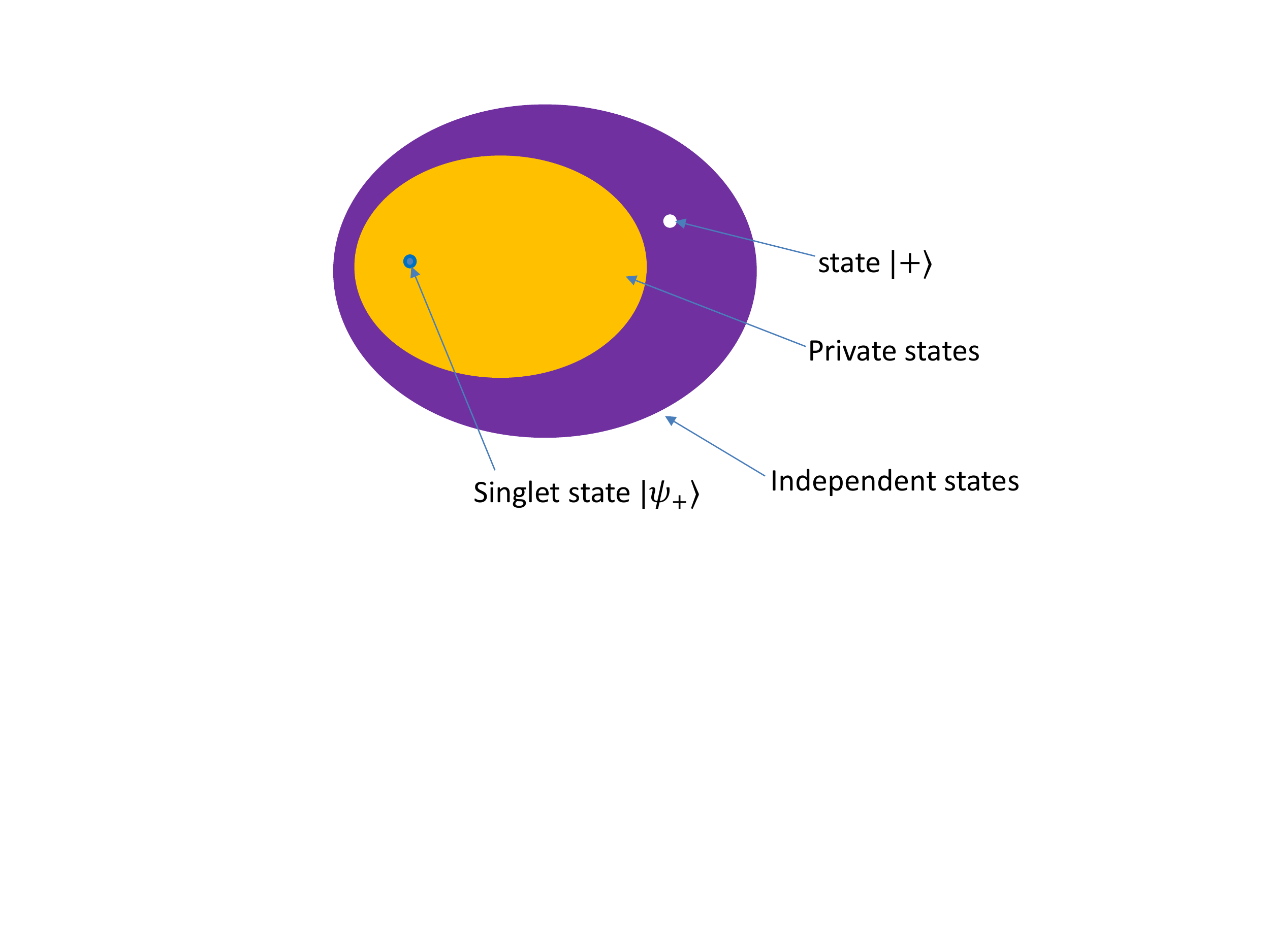}
    \caption{The onion structure of quantum states containing ideal privacy. The singlet state is an example of a private state. The set of private states is a proper subset of the set of independent states. The state $|+\rangle$ is an independent state, which is not a private state.}
    \label{fig:onion}
\end{figure}

It is common in the literature to represent the private randomness obtained by two honest parties against an eavesdropper in terms of {\it tripartite} states $\left(\sum_{i=0}^{d_A-1}{1\over d_A} |i\>\<i|)\otimes
\sum_{k=0}^{d_B-1}{1\over d_B} |k\>\<k|\right)\otimes\rho_{E}$. (here $\rho_{E}$ is representing an arbitrary state of the eavesdropper) In such approach the honest parties are using {\it local operations and public communication}. However it is shown in \cite{YHW}, that this approach is equivalent to distilling specific {\it bipartite} states - the independent states, by means of CLODCC operations.
The independent states have form
of coherence "twisted" into a shared mixed state:
\begin{equation}
    \alpha_{d_A,d_B} = \sum_{i,j,k,l}
    |i\>\<j|\otimes |k\>\<l| \otimes U_{ik} \sigma_{A'B'} U_{jl}^{\dagger},
\end{equation}
as it can be written in the following way:
\begin{equation}
    \tau |+\>\<+|_A \otimes |+\>\<+|_B \otimes \sigma_{A'B'}\tau^{\dagger}
\end{equation}
 with $|+\>_{A/B} = \sum_{i=0}^{d_{A/B}-1}{1\over \sqrt{d_{A/B}}} |i\>$ and $\tau = \sum_{ij} |ij\>\<ij|\otimes U_{ij}^{A'B'}$.

 In the scenario 
 considered here, i.e when two parties want to localize private randomnes at one place, we will be interested
 in {\it local} independent states:
 
 \begin{equation}
    \alpha_{d_A} = \sum_{i,j}
    |i\>\<j| \otimes U_{i} \sigma_{A'B'} U_{j}^{\dagger}
\end{equation}
 
 \begin{figure}[h!]
 \begin{centering}
\includegraphics[trim= 0cm 0cm 0cm 0cm, width=1.4\columnwidth]{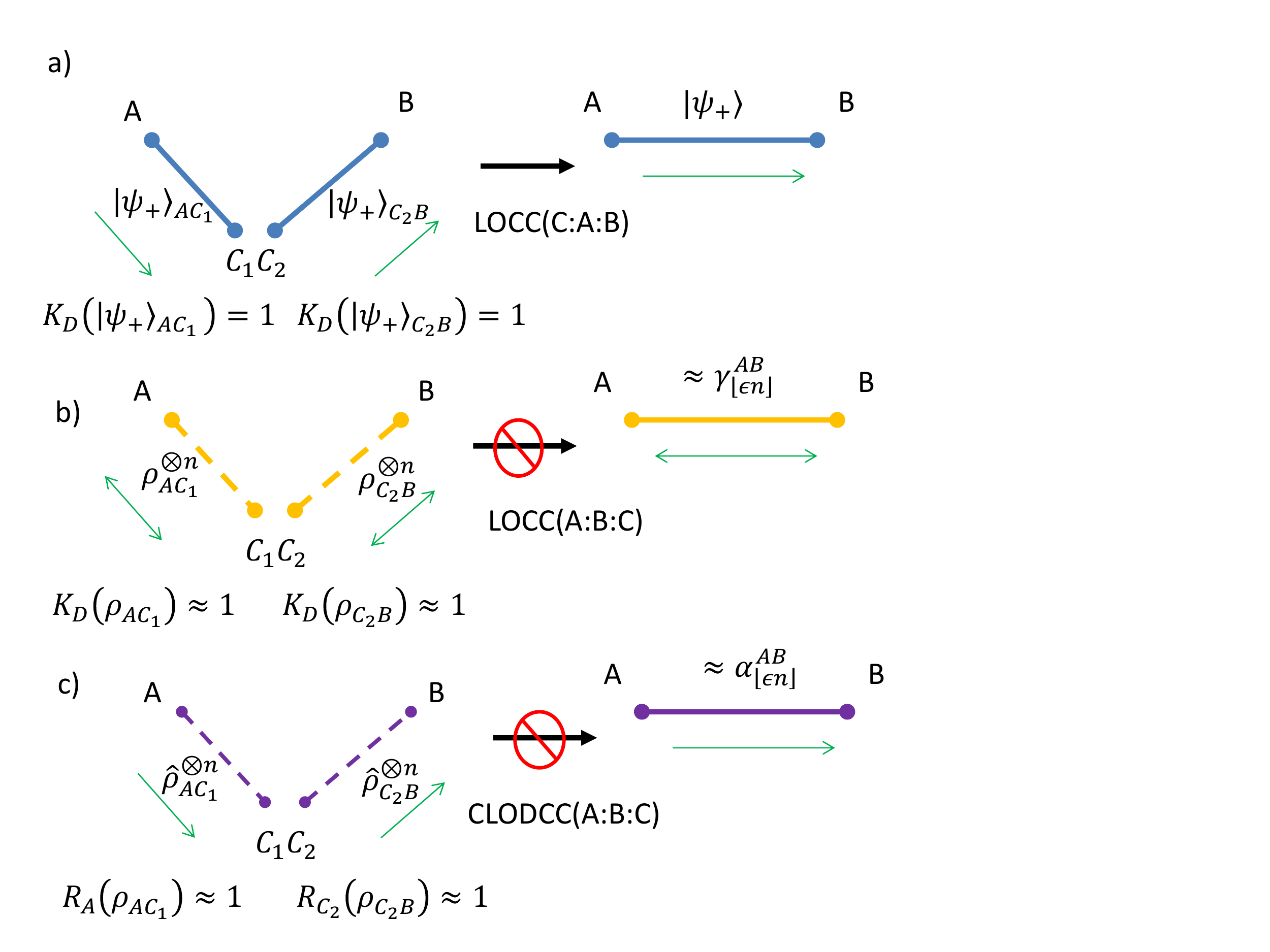}
\par\end{centering}
 \caption{\label{fig:main} Depiction of the limitation for private randomness in the context of network repeaters. Part a) depicts redistribution of loyalty in the network via entanglement swapping: on the LHS $A$ trusts $C$ and $C$ trusts $B$, as depicted by the green arrows. On the RHS $A$ trusts $B$ only \cite{ent_swapping}. Part b) shows that for any $\epsilon>0$ there exist states with positive partial transposition, that have almost $1$ bit of secure key $K_D$ each. However, there is no LOCC protocol between three parties that outputs an approximate private state with more than $\lfloor n\epsilon \rfloor$ bits of key \cite{BCHW}. Part c) depicts the result of this paper, in analogy to the case b): for any $\epsilon>0$, there exist states that have almost $1$ bit of private randomness, but there is no $\mathrm{CLODCC}$ protocol between three parties that outputs {\it an independent state} with more than $\lfloor n\epsilon \rfloor$ bits of private randomness.}
 \end{figure}
Although the structural analogy between the theories of private key and randomness is somewhat natural, the results explicitly determining this relation are missing. Developing this analogy, we first show that any state containing ideal private key (a private state) \cite{pptkey,keyhuge} is, in fact, an independent state.  We therefore prove that the sets of quantum states containing ideal privacy form an onion structure (see Fig. \ref{fig:onion}).

We then demonstrate that private randomness exhibits the similar type of limitation as a secure key when distributed on a communication network  \cite{BCHW}. The answer to the question ``Can one always swap private randomness of general mixed quantum states?'' follows this close analogy. 

The conceptual description that we introduce to capture the topology of security in the network is called the {\it loyalty network}. It represents each party as a vertex, while a directed edge from vertex $A$ to vertex $B$ represents $A$ being secure due to loyalty of $B$. In the weaker sense, loyalty $A \rightarrow B$  means that $A$ trusts that $B$ will not hand over his subsystem $\rho_B$ of the shared joint state $\rho_{AB}$ to any eavesdropper Eve.  Clearly, if $B$ is not loyal to $A$,  the local private randomness of $A$ is equal to localisable purity. However, we will assume a stronger sense of loyalty, in which loyal $B$ cooperates in favor of $A$, such that $A$ has access to as much of private randomness of a state $\rho_{AB}$ as it is possible  (part of it is obtained from the correlations between $A$ and $B$). 

We will exemplify this concept with entanglement swapping of the singlet $|\psi\>_+^{AB}:={1\over {\sqrt{2}}}(|00\>+|11\>)_{AB}$:
\begin{equation}
|\psi\>_{+}^{AC_1}\otimes |\psi\>_{+}^{C_2B}\xrightarrow{\mbox{ent. swap.}} |\psi\>^{AB}_{+}.
\label{eq:ln}
\end{equation}
This operation can be interpreted as:
\begin{itemize}
    \item {\it Initially, party $A$ has $1$ bit of private randomness due to loyalty of party $C = (C_1C_2)$, and party $C$ has $1$ bit of private randomness due to loyalty of party $B$. After applying entanglement swapping, party $A$ has $1$ bit of private randomness due to loyalty of party $B$, and does not need to rely on the loyalty of party $C$ anymore.}\footnote{Another way to see the dependencies in the loyalty network of \eqref{eq:ln} is: at the beginning $A$ trusts $C$ and $B$ trusts $C$, while the task is to remove $C$ out of the network and to make $A$ trusting $B$ (or $B$ trusting $A$). As we show in Section \ref{sec:main-results}, the bound  for repeated private randomness which we provide is invariant under the swap, hence it covers also this particular topology of network. Moreover, in Section \ref{sec:examples} we show that there are swap-invariant states (e.g., some Werner states) that exhibit gap between localisable and repeated private randomness.}
\end{itemize} 
We then ask if such transformation is possible for all mixed quantum states, when the number $n$ of copies of initial states goes to infinity,
\begin{equation}
(\rho_{AC_1}\otimes {\tilde{\rho}}_{C_2B})^{\otimes n}\xrightarrow[n\rightarrow \infty]{\mbox{priv. rand. repeater ?}} \alpha^{AB}_{k\times n},
\end{equation}
where $k\times n$ is the rate of private randomness that can be obtained via tripartite operations from $n$ copies  of the input state in the form of the {\it independent states}. These states, denoted by $\alpha$, contain ideal private randomness directly accessible by local complete von Neumann measurement on subsystem of $\alpha$. For the qualitative summary of the results, see Fig. \ref{fig:main} and Section~\ref{sec:summary}.

Since we adopt methods shown in
\cite{BCHW}, the upper bound that we obtain
works for the states with {\it positive partial transposition} (PPT states). These are bipartite states $\rho$ that satisfy $(\id\otimes(\cdot)^\top) (\rho) \geq 0$ \cite{Peres-book}, where $(\cdot)^\top$ is a transposition and $\id$ is an identity operator. We show the power of the upper bound by inspecting the gap between localisable and repeated private randomness for separable Werner states. These are states interpolating between {\it symmetric} and {\it antisymmetric} state. Within the range of interpolating parameter that guarantees separability, for sufficiently large local dimension $d$, we observe the presence of a gap. We also consider a strictly smaller class of operations, generated by compositions of: (i) $n$ optimal single copy operations among the three parties, followed by (ii) distillation by $A$ and $B$ solely, via general $\mathrm{CLODCC}(A:B)$ operations. For this class, we derive bound for repeated private randomness for arbitrary states. We then exemplify it by providing a family of states that do not have positive partial transposition, yet exhibit the same gap (of almost $1$) between localisable and repeated randomness.

\subsection{Summary of the main results}
\label{sec:summary}
For reader's convenience, we summarise here the main results of our contribution. 

Here and further in this paper we write interchangeably $\rho_{AC_1},\tilde{\rho}_{C_2B}$ and $\rho,\tilde{\rho}$ whenever it is clear from the context. Given a state $\rho_{AB}$, $S(A)_\rho$ will denote the von Neumann entropy of subsystem $A$ of $\rho_{AB}$, $S(A)_\rho:=-\tr(\rho_A\log \rho_A)$ with $\rho_A=\tr_B \rho_{AB}$. By $S(A|B)_\rho$  we denote the conditional entropy $S(AB)_\rho- S(A)_\rho$, while $I(A:B)_\rho = S(A)_\rho +S(B)_\rho -S(AB)_\rho$ is the quantum mutual information. By $\log |A|$ we mean the $\log$ of dimension of the system $A$ (similarly for $B$ and $AB$). In case it is necessary, we will explicitly write the state of which dimension is invoked: $\log |A|_\rho$. The logarithm is of the base $2$ throughout all of this paper. For special case of a distribution $\{p,(1-p)\}$, its Shannon entropy we denote as $h(p)= - p\log p - (1-p)\log (1-p)$. We also refer to it as to {\it binary Shannon entropy}.

The main step towards our results is the definition of the {\it repeated randomness} ${\cal R}_A^{A\leftrightarrow C_1C_2 \leftrightarrow B} (\rho\otimes {\tilde{\rho}})$ (an analog of repeated key), which is the asymptotic rate ($n\rightarrow\infty$) of private randomness (in the form of the ibits), that can be achieved by the three parties from initial $n$ copies of the state $\rho\otimes {\tilde{\rho}}$ via operations allowed in the resource theory of private randomness \cite{YHW} (called $\mathrm{CLODCC}$). Separately, we define private randomness repeater rate in the i.i.d. case, i.e., when three parties perform the same  $\mathrm{CLODCC}$ operation on each of the copies of the state, followed by a general operation from $\mathrm{CLODCC}(A:B)$.

As the main result,
we prove the following upper bound on the rate of repeated private randomness:
\begin{align}
{\cal R}_A^{A\leftrightarrow C_1C_2 \leftrightarrow B} (\rho_{AC_1}\otimes {\tilde{\rho}}_{C_2B}) &\leq& \nonumber\\ D(\rho^{\Gamma}||{\id\over |AC_1|})+ 
D({\tilde\rho}^{\Gamma}||{\id\over |C_2B|}),
\label{upp.bound.rpr}
\end{align}
where $D(\rho||\sigma)=\tr(\rho\log \rho-\rho\log\sigma)$ is the quantum relative entropy, 
and $\rho^\Gamma:=(\id\otimes T)(\rho)$ denotes the {\it partial transposition} of $\rho$. The RHS of \eqref{upp.bound.rpr} can be quite small in some cases, as we show with particular examples of states for which repeated private randomness is negligible. 
It can be rephrased in terms of the {\it global purity}, $G(\rho_{XY}) := \log|XY|-S(XY)_\rho$, as
\begin{equation}
{\cal R}_A^{A\leftrightarrow C_1C_2 \leftrightarrow B}(\rho_{AC_1}\otimes \rho_{C_2B}) \leq G(\rho_{AC_1}^{\Gamma})+G({\tilde \rho}_{C_2B}^{\Gamma}).
\label{eq:corollary1}
\end{equation} 
The above form of the bound would be natural in purity distillation paradigm \cite{huge-delta}. In the context of private randomness distillation it will be also natural to rephrase it in terms of correlations i.e. quantum mutual information. This is because the mutual information quantifies the nontrivial (not equivalent to purity) amount of private randomness. For the case when $\rho={\tilde \rho}$ has positive partial transposition and has both subsystems in maximally mixed states,
we have immediate corollary \footnote{The partial transposition does not change the entropy of neither of the subsystems of $\rho$. One subsystem (say $A$) is the same after applying the map $\id_A\otimes(\cdot)^\top_B$. For the other, by the fact that $\mathrm{det}(X)=\mathrm{det}(X^\top)$ one has $\mathrm{det}(\rho_B - \lambda\id)=\mathrm{det}(\rho_B - \lambda\id)^\top=\mathrm{det}(\rho_B^\top - \lambda\id)$. Hence, the roots of this polynomial, which are the eigenvalues of $\rho_{B}$, are the same as for $\rho_B^\top$.}:
\begin{equation}
{\cal R}_A^{A\leftrightarrow C_1C_2 \leftrightarrow B}(\rho_{AC_1}\otimes \rho_{C_2B}) \leq 2I(A:C_1)_{\rho^{\Gamma}}.
\label{eq:corollary}
\end{equation}
This stems from the fact, that the quantum relative entropy between a state and the product of its two subsystems is equal to quantum mutual information between them.

The key result of \cite{YHW} which allows us to interpret our main result is the protocol of optimal private randomness distillation. It determines how a single party can localise as much of private randomness in her system as possible. Additionally, in \cite{YHW} it is shown that there are two sources of private randomness: local, in form of purity, and shared, in form
of correlations. This fact is supported by quantitative result: the amount of localised private randomness of a state with positive partial transposition $\rho_{AC_1}$ in the asymptotic limit reads
\begin{equation}
\begin{split}
R_A (\rho_{AC_1}) =&\;\; (\log |A| - S(A)_\rho) + (\log|C_1| -S(C_1)_\rho)\\ &\;\;+ I(A:C_1)_\rho.
\end{split}
\end{equation}
Thus, the amount of locally achievable private randomness for the $\rho_{AC_1}$ (i.e., between $A$ and $C_1$) equals to the sum of local purity $\log |A| -S(A)_\rho$ and the amount of correlation in the shared state (i.e., the quantum mutual information). When the state $\rho$ has subsystems in maximally mixed state, we can use the bound from Eq. (\ref{eq:corollary}) since no local purity can be achieved, i.e. $R_A(\rho_{AC_1})=I(A:C_1)_{\rho}$. It applies  for states with positive partial transposition, for which there is a gap:
\begin{equation}
I(A:C_1)_\rho > 2 I(A:C_1)_{\rho^{\Gamma}}.
\end{equation}
Although we notice the gap between correlations of $\rho$ and $\rho^{\Gamma}$  for states having key (and therefore distillable private randomness) \cite{smallkey}, the above gap can not be demonstrated in the same way
as in \cite{BCHW} due to the factor $2$ above. Instead, since the key is not the only local form of private randomness, we study the most famous single parameter class of states, the Werner states. In particular, we observe that the {\it symmetric} Werner state \cite{Werner1989}, $\rho_s^d= {1\over d^2 +d}\left(\id +V\right)$,  where $V:=\sum_{i,j}|ij\>\< ji|$ is called a \textit{swap} operator, satisfies
\begin{equation}
\begin{split}
&I(A:B)_{\rho_s^d} = 1 + \log \left( {d \over d+1}\right) \xrightarrow[d\rightarrow \infty]\, 1,\nonumber\\
&I(A:B)_{(\rho_s^d)^{\Gamma}} = {1\over d} \log d + {d-1\over d} \log {d \over d-1} \xrightarrow[d\rightarrow \infty]\, 0.
\end{split}
\end{equation}
Hence, for large dimensions of $d$,
\begin{eqnarray}
R_A(\rho^d_s) \approx 1 \quad \text{and} \quad {\cal R}_A(\rho^d_s\otimes \rho^d_s) \approx 0.
\end{eqnarray}
As we show in Section \ref{sec:examples}, any separable Werner state of sufficiently high dimension exhibits the gap, as it is the case for the symmetric one. This result is analogous to the limitation for key repeaters shown in \cite{BCHW}. In contrast, however, it is achieved on separable states, rather than on the approximate private states used in \cite{BCHW}.

Finally, we consider a variant of  i.i.d. case, when the three parties are forced to use identical operations on each copy of the state and, further, $A$ and $B$ apply any $\mathrm{CLODCC}(A:B)$ on such obtained outputs. For a particular independent state of the form
\be
\alpha_{V,d} = {1\over 2}\left[\bea{cc}
{\id\over d^2} &{V \over d^2} \\ \\
{V \over d^2}& {\id\over d^2} \\
\eea
\right],
\label{eq:alpha1}
\ee
we prove the existence of a gap between private randomness $R_A(\alpha_{V,d}), R_B(\alpha_{V,d})$ and i.i.d. repeated private randomness $\mathcal{R}_A^{\operatorname{iid}}(\alpha_{V,d})$, whenever dimension is sufficiently large. Namely, we prove that, for $d>32$, we have  $R_A(\alpha_{V,d})= R_B(\alpha_{V,d})= 1$, while $\mathcal{R}_A^{\operatorname{iid}}(\alpha_{V,d})<1$. In particular, for $d>11$, we have
\begin{equation}
            \mathcal{R}_A^{\operatorname{iid}}(\alpha_{V,d})\leq {{4\log d}\over d} + \eta({{4}\over d}),
        \end{equation}
which clearly goes to 0 when $d\rightarrow \infty$.

Our paper is organized as follows. We start from Section~\ref{preliminaries}, where all necessary tools are presented; in particular, we introduce the concepts of the $\mathrm{CLODCC}$ operations and of a local idit. In Subsection~\ref{sec:scenario} we precisely describe the framework in which we work, stating what the involved parties are allowed to perform. We do so by defining the allowed class of operations ($\mathrm{CLODCC}$) and its distinguished subclass, and by establishing relations between them. 
Section~\ref{sec:ps-are-is} analyses the relationship between the sets of private states and of independent states, showing they are not equal to each other. This finding certifies the novelty of our work in comparison to previous results on limitations on quantum key repeaters.  Section~\ref{sec:main-results} is divided into two separate parts. In the first one, we prove new results on the state discrimination from the maximal noise by using $\mathrm{CLODCC}$ operations. In the second part, we derive our main result: an upper bound on the rate of repeated randomness. This implies existence of the states with localisable randomness equal $1$ that have vanishingly small repeated independent randomness.  In Section~\ref{sec:direct} we provide alternative proof of the bound on repeated private randomness for states with positive partial transposition, showing, as a byproduct, that the latter rate is bounded by a value computed on partially transposed states.
In Section~\ref{sec:limitations_iid}, we present the limitation for private randomness repeater in the i.i.d. case, where parties first perform the same $\mathrm{CLODCC}$ operation on each copy of the state and then apply arbitrary $\mathrm{CLODCC}$ on these copies. In particular, for a chosen class of independent states and for sufficiently large dimension, we show a gap between the private randomness and the repeated private randomness. In Section~\ref{sec:examples} we show a broad class of Werner states for which our main result holds. We close this paper with Section~\ref{sec:discussion}, summarising our main results and putting them in the broader picture of possible further research.

\section{Preliminaries on private randomness and key}
\label{preliminaries}
In this Section, we recall necessary concepts of the resource theory of private randomness and private key, allowing the reader to better understand our further results.

The free operations of this theory are {\it closed operations and classical communication via dephasing channel} ($\mathrm{CLODCC}$). This class of operations is a subclass of the well known $\mathrm{LOCC}$ operations, and was introduced as free operations in the resource theory of purity \cite{OHHH2001}. The systems under consideration are closed, only local unitary transformations are allowed, and the honest parties can exchange subsystems through a dephasing channel. Such dephasing channel can be realized by an eavesdropper Eve via: (1) attaching and ancillary pure state $|0\>_E$ to each system $M$ passing between the honest parties, (2) performing a CNOT gate (with source at $M$ and target at the system $E$), and (3) collecting $E$ in some quantum memory. 

The {\it target states} (i.e., states containing ideal private randomness in a directly accessible form) are given by {\it independent states} \cite{YHW}, which can be viewed as the result of  {\it twisting} of coherent states \cite{Alex-Review} ${1\over {\sqrt{d}}}(\sum_{i=0}^{d-1}|i\>_A)\otimes {1\over {\sqrt{d}}}(\sum_{i=0}^{d-1}|i\>_B)$. In case of two dits of private randomness the independent states have the form
 \begin{equation}
 \alpha_{ABA'B'}:= U |+\>\<+|_A\otimes|+\>\<+|_B \otimes\sigma_{A'B'} U^{\dagger},
 \label{eq:ibits}
 \end{equation}
 where $U= \sum_{i,j} |ij\>\<ij|\otimes U_{ij}$ and $U_{ij}$ is a unitary transformation for each $ij$. 
 
By {\it local idit} we will mean the independent state given in Eq. (\ref{eq:ibits}) when $|A|=d$ and $|B|=1$ (or $|A|=1$ and $|B|=d$). Hence, private randomness can be directly accessed from a part of such state that is localised either at $A$ or at $B$.  To explicitly indicate the number $m$ of private random bits directly accessible via measuring systems $A$ (or $B$) in a local idit, we will denote it as $\alpha_m$.
 
Note that these states are similar in construction to the private states, defined \cite{pptkey,keyhuge} by {\it twisting} of maximally entangled states $|\psi\>_+^{AB}:={1\over {\sqrt{d}}}\sum_{i=0}^{d-1}|ii\>_{AB}$,
\begin{equation}
\gamma_{ABA'B'}:= U |\psi\>_+\<\psi|_{AB}\otimes \sigma_{A'B'} U^{\dagger},
\end{equation}
with the unitary operator $U=\sum_{i} |ii\>\<ii|\otimes U_i$. Every key distillation protocol ends up in states approximating private states, while every protocol which distills private randomness produces approximated independent states.
In Section \ref{sec:ps-are-is} we show that any private state is an independent state.

Following \cite{YHW}, $R_A(\rho)$ will denote the {\it private randomness localisable on system $A$ by means of $\mathrm{CLODCC}(A:B)$ operations from (asymptotically many) copies of $\rho_{AB}$}.

An important result from
\cite{YHW} asserts: if a bipartite state has a negative conditional entropy, then the whole of its private randomness content can be localised at each of parties by means of $\mathrm{CLODCC}$ operations:
\begin{theorem} [Corollary from Theorem $4$ of \cite{YHW}]
Any bipartite state $\rho_{AB}$ satisfying $S(B|A)_\rho > 0$ satisfies:
$R_{A}(\rho_{AB}) = \log|AB| - S(AB)_\rho$.
\label{thm:neg-coh-random}
\end{theorem}
The quantity $\log|AB| - S(AB)_\rho$ is called a {\it global purity} \cite{OHHH2001}, and is also a trivial upper bound on the amount of localisable private randomness (achieved when both parties can operate globally on the system $AB$). Any separable state and, in general, states with positive partial transposition have positive quantum conditional entropy (i.e., negative {\it coherent information}) \cite{RMPK-quant-ent}. Moreover, as we will see, some ibits that have negative partial transposition share this property with PPT states. Furthermore, the resource theory of private randomness has an empty set of free states: adding a maximally mixed state can increase the amount of localisable private randomness. However, the maximally mixed state on its own represents the set of states which are closed under $\mathrm{CLODCC}$ operations (see Section \ref{sec:main-results}) and it contains zero localisable private randomness. We can, therefore, view this state as a correspondent
of the set of separable states in the resource theory of private key.

In what follows  $\rho \approx_\epsilon \rho'$ denotes $||\rho - \rho'||_1\leq \epsilon$ with $||X||_1:=\tr{|X|}$ for a hermitian operator $X$.

\subsection{The scenario of private randomness repeaters}
\label{sec:scenario}
\begin{figure}[h!]
 \begin{centering}
\includegraphics[trim= 0cm 8cm 0cm 0cm, width=1.15\columnwidth]{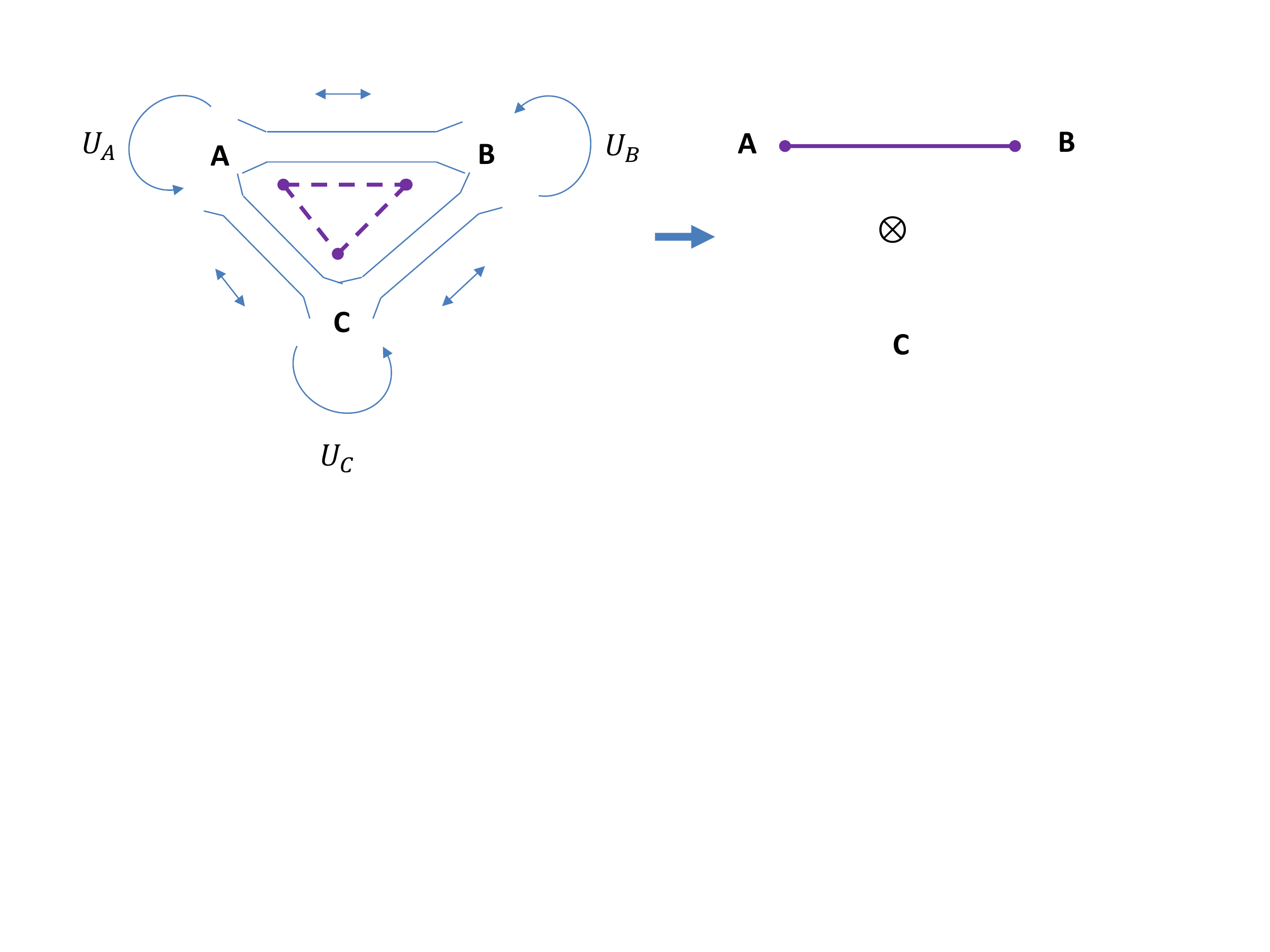}
\par\end{centering}
 \caption{\label{fig:scen} Depiction of the considered scenario. All the three parties can perform locally unitary transformations, and can send a system down a dephasing channel to the other parties. Their task is to distill {\it independent states} shared by $A$ and $B$.}
 \end{figure}

In our scenario, there are three involved parties: $A$, $B$, and $C$. Party $C$ has two subsystems: $C_1$ and $C_2$. A dephasing channel connects each pair of parties. Each of the parties can perform either  (i) unitary operation, or (ii) sending of a system to some
of the other parties (or both of them). We denote as $\mathrm{CLODCC}(A:C_1C_2:B)$ the class of operation generated by arbitrary (possibly infinite) compositions of the above operations. The parties are given (arbitrarily large) $n$ copies
of input states $\rho_{AC_1}$ and $\rho_{C_2B}$ shared by 
$A$ and $C$, and $C$ and $B$, respectively.
The task of the parties is to obtain a local idit $\alpha_m$ on systems $A$ and $B$ with the largest possible amount $m$ of bits of private randomness, with randomness directly accessible by
von Neumann measurement on Alice's system (see part c) of Figure \ref{fig:main} and Figure \ref{fig:scen}). In the case of the above scenario, we obtain the bound for states with positive partial transposition.

To obtain analogous results for states that are not having positive partial transposition, we will consider a much simpler scenario, with smaller class of allowed operations, $\mathrm{CLODCC}(C^{\mbox{iid}}:(A^{\mbox{ iid}}:B^{\mbox{iid}})) \subsetneq \mathrm{CLODCC}(A:B:C)$. For the case of $n$ copies of 
the input states, this class is defined by the composition of  two operations, denoted as $(C:A:B)^{\mbox{iid}}$ and $A\leftrightarrow B$, respectively. The former operation corresponds to an action of the three parties: they behave identically on each copy, producing $n$ copies of the best single-round output $\hat{\rho}$. The latter operation refers to $A$ and $B$ performing general $\mathrm{CLODCC}(A:B)$ operation on $\hat{\rho}^{\otimes n}$. 
The task for the parties is again to distill independent states shared by $A$ and $B$.

We end this Section with several simple observations, which are crucial for our later considerations.
\begin{observation}
There is $\mathrm{CLODCC}(A:C_1C_2:B) \subset \mathrm{CLODCC}(C_1C_2 : AB)$.
\label{obs:CLODCCinclusions}
\end{observation}
{\it Proof}. The difference between these two sets of
operations is that $A$ and $B$ are joining
their labs. They can now perform global unitary transformations, and we have to show that they still are
be able to dephase parts of their system.
E.g., $A$ can send a state to $B$ via dephasing channel according to definition of the set of $\mathrm{CLODCC}(A:C_1C_2:B)$ (and vice versa). When $A$ and $B$ acting together want to dephase some system, they can send it to $C$ who sends it back to them. The claim is then seen from the fact that single dephasing channel between $C$ and $AB$ can
also simulate two separate dephasing channels
between $C$ and $A$, and $C$ and $B$ respectively, while operations of $C$ are the same in both sets.
$\square$

Consider the set $S$ of operations on system $AB$ induced from the operations in $\mathrm{CLODCC}(C:AB)$ via composing the latter with a partial trace over system $C$. We will argue that this set includes operations that are composition of unitary transformations and projections in computational basis. We will denote the set of all such compositions as $\mathrm{U+Deph}$.

\begin{observation}  The set $S$ of transformations of system $AB$ defined as $\tr_C \Lambda_{C:AB}(\rho_{ABC})$, with $\Lambda_{C:AB}\in \mathrm{CLODCC}(C:AB)$, satisfies $\mathrm{U+Deph} \subset S$.
\label{obs:induced}
\end{observation}
 {\it Proof}. 
 It follows directly from the fact that operations $U_{AB}\otimes \id_C$ and $\{P_{a}\otimes \id_C\}$ with $a$ being subsystem of $AB$, belong to the set $\mathrm{CLODCC}(C:AB)$. Indeed, the von Neumann measurement on subsystem of $AB$ can be realized via composition of sending the measured system $a$ to $C$ and re-sending it back to $AB$. The same holds for arbitrary composition of the latter two. The assertion then follows 
from the fact that $\tr_C(L_{AB}\otimes \id_C )(\rho_{ABC})=
L_{AB}(\tr_C\rho_{ABC})$ for any completely positive trace preserving linear map $L_{AB}$.
$\square$

It is common that the allowed operations in a given 
resource theory preserve the set of the free states, i.e.,
transform any free state into a free state. The observation
below implements this property for the 
resource theory of (distributed) private randomness.

\begin{observation}
\label{obs1}
Every $\Lambda \in \mathrm{CLODCC}$ is unital, i.e., $\Lambda$ preserves the maximally mixed state.
\end{observation}

{\it Proof}.
According to the definition of $\mathrm{CLODCC}$, presented in Section~\ref{preliminaries}, operations in this class are composed of unitary operations and dephasing together with sending dephased system from one party to another.
Clearly, the first two operations preserve the maximally mixed state. The only nonunital operation is sending of the dephased system. However, a subsystem of a maximally mixed state is also a maximally mixed one, hence the map outputs also a maximally mixed state, but (possibly) of different dimension on systems $A$, $B$ and $C$ (denoted as $|\hat{A}|$, $|\hat{B}|$ and $|\hat{C}|$, respectively). However, $|A|+|B|+|C| = |\hat{A}|+|\hat{B}| +|\hat{C}|$, because CLODCC class does not contain the partial trace operation. Hence, this map can be seen as ``redistributing'' the maximally mixed 
state among the three systems.
$\square$

\section{Private states are independent states}
\label{sec:ps-are-is}
In this Section, we discuss the differences between private states and independent states. In particular, we prove that the set of independent states is strictly included in the set of all private states. This follows from the fact that there are product states, such as $|+\>\otimes \id /2$, which are ibits having zero distillable key, because entanglement is a precondition for secure key \cite{CurtyLewLut}. Nevertheless, the techniques used here are related to those in \cite{BCHW}. For example, the relative entropy is taken with respect to the set of separable states, while here it is taken with respect to the maximally mixed state. We have to simplify the approach, because the private randomness is zero for the maximally mixed state, and is non-zero for any other state. We show that these two similar, although different classes of states, are related by the strict inclusion $PS \subsetneq IS$, which is the main result of this Section.

{\proposition Any private state is a (local) independent state, while the converse statement is not valid in general, $PS \subsetneq IS$. Moreover, the private random bit can be located at either of the
parties.}

{\it Proof}. Any private state has a form $\gamma_{ABA'B'} = \sum_{i,j=0}^{d-1}{1\over d} |ii\>\<jj|\otimes U_i\sigma_{A'B'} U_j^{\dagger}$. The twisting involved in the definition of any private state can be simplified to have a single control \cite{bigkey}: 
\begin{widetext}
\begin{equation} 
\label{single_control}
\gamma_{ABA'B'}= \left( \sum_{i=0}^{d-1}|i\>\<i|_A\otimes\id_B\otimes U_i\right) |\psi_+\>\<\psi_+|_{AB} \otimes\sigma_{A'B'} \left(\sum_{j=0}^{d-1} |j\>\<j|_A\otimes \id_B \otimes U_j^{\dagger}\right).
\end{equation}
\end{widetext}
It is then enough to express the singlet state $|\psi_+\>_{AB}$ as an output of a control-shift gate:
$|\psi_+\>_{AB} =\tau |+\>_A\otimes |0\>_B$ with $|+\> = \sum_{i=0}^{d-1}{1\over \sqrt{d}}|i\>$ and
$\tau = \sum_i |i\>\<i|_A\otimes S_{i,d}$, where $S_{i,d} |j\>  = |j+i \,\mbox{mod}\, d\>$, if $d$ is prime. If $d$ is not prime, it can be expressed uniquely by multiplication of primes: $d=d_1\times \cdots \times d_k$ where $d_l$ is prime for $l\in\{1,\ldots,k\}$ (for the sake of uniqueness, we assume $d_l \leq d_{l'}$ for $l\leq l'$). In this case we define $\tau := \bigotimes_{l=1}^{k} \left(\sum_{i=0}^{d_l-1} |i\>\<i|\otimes S_{i,d_l}\right)$, where $S_{i,d_l}$ is defined as above with $d_l$ in place of $d$. Substituting this form of a private state into \eqref{single_control} immediately yields
\begin{widetext}
\begin{equation}
\gamma_{ABA'B'}= \left(\sum_{i=0}^{d-1} |i\>\<i|_A \otimes \left(\bigotimes_{l=1}^{k} S_{l[i],d_l}\right)\otimes U_i\right) |+\>\<+|_{A}\otimes |0\>\<0|_B \otimes\sigma_{A'B'} \left(\sum_{j=0}^{d-1} |j\>\<j|_A\otimes \left(\bigotimes_{l=1}^{k} S_{l[j],d_l}^{\dagger}\right) \otimes U_j^{\dagger}\right),
\end{equation}
\end{widetext}
where $l[i]$ is the $l$-th digit of $i$
written in a multi-base system of $k$ bases:
$d_1,\ldots,d_k$. Written in such a form, this state is by definition a (local) independent state. Indeed, consider Eq. (\ref{eq:ibits}), with substitution $B$ of system of dimension $1$ and $B'$ system in state $|0\>\<0|\otimes \tr_{A'}\sigma$.
The strictness of inclusion follows from the state $|+\>\otimes {\id/2}$ being an ibit, while having no distillable key, because entanglement is a precondition of security \cite{CurtyLewLut}. Because the singlet state is swap invariant, the same reasoning follows when one expresses it as 
$|\psi_+\>_{AB} =\tau' |+\>_B\otimes |0\>_A$ with $\tau'$ having control at $B$ rather than at $A$. This fact shows that the private random bit can be located in any of the parties. \qed

The above Theorem implies the onion structure of quantum states containing ideal privacy:
$|\psi_+\> \in PS \subsetneq IS$ and $|+\> \in IS \setminus PS$ (see Fig. \ref{fig:onion}).

\section{Limitations on private randomness repeaters}
\label{sec:main-results}
The main result of this Section provides a bound on repeated independent randomness. It is based on restricted relative entropy bound of the supplemental material of \cite{BCHW}, with the difference that 
allowed operations are taken to be $\mathrm{CLODCC}$ instead of $\mathrm{LOCC}$,
while the set of free states is given by a maximally mixed state, instead of the set of separable states. We will first describe the asymptotic distinguishability using operations from $\mathrm{CLODCC}$. 

\subsection{Discriminating states from maximal noise via $\mathrm{CLODCC}$ operations}
\label{sec:dist-bound}

We are interested in an asymptotic distinguishability. In analogy to restricted relative entropy of entanglement of \cite{Piani2009-relent}, we consider now the simplest of the restricted relative entropy: the relative entropy with respect to the maximally mixed state. Due to limitations of the specific technique, our results hold only for states with positive partial transposition (PPT states). We build on the results of \cite{BCHW}.

{\definition For a bipartite state on $\mathcal{H}:=\mathbb{C}^d\ot\mathbb{C}^d$, the restricted relative entropy distance from maximal mixed state achievable via operations from set a $S$ of POVMS is
\begin{align}
D_S(\rho)&:=\sup_{\Lambda_M\in S} D\left(\Lambda_M(\rho)||\Lambda_M\left({\mathbb{I}\over d^2}\right)\right),\label{DS.definition}\\
D^{\infty}_S(\rho)&:=\lim_{n\rightarrow\infty}{1\over n} D_S(\rho^{\ot n}),\label{DS.infty.definition}
\end{align}
where $\Lambda_M:=\sum_i\tr_\mathcal{H}(M^i(\cdot))|i\rangle\langle i|$ is a completely positive trace-preserving map, $n\in\mathbb{N}$, and $D(\cdot||\cdot)$ is the Kullback--Leibler relative entropy of two probability distributions. A restriction of $S$ in \eqref{DS.definition} and \eqref{DS.infty.definition} to the set $J$, corresponding to such $\Lambda_M$ that belong to $\mathrm{CLODCC}$ class, defines $D_J(\rho)$ and $D_J^\infty(\rho)$, respectively.}

{\theorem If $\rho$ is a density operator on $\mathcal{H}:=\mathbb{C}^d\ot\mathbb{C}^d$, $\Gamma:=\mathrm{id}_{\mathbb{C}^d}\otimes(\cdot)^\top$, and $X^\Gamma:=\Gamma(X)$ for a linear bounded $X:\mathcal{H}\rightarrow\mathcal{H}$, then 
\be
\rho^\Gamma\geq0\;\;\Rightarrow\;\; D_J^{\infty}(\rho)\leq D(\rho^{\Gamma}||{\mathbb{I}\over d^2}).
\ee
\label{eq:dist-thm} 
}

{\it Proof}. Let $\Lambda\in\mathrm{CLODCC}$ and let $\{|i\rangle\langle i|\}$ be a base in $\mathcal{H}$. Then
\begin{widetext}
\begin{equation}
\begin{split}
    &\sup_{\Lambda\in\mathrm{CLODCC}}D\left(\Lambda(\rho^{\otimes n})||\Lambda\left(\frac{\mathbb{I}^{\otimes n}}{d^{2n}}\right)\right)\\
    &:=\sup_{\Lambda\in\mathrm{CLODCC}}D\left(\sum_i\tr_\mathcal{H}(M_\Lambda^i\rho^{\otimes n})\otimes|i\rangle\langle i|||\sum_i\tr_\mathcal{H}\left(M^i_\Lambda\frac{\mathbb{I}^{\otimes n}}{d^{2n}}\right)\otimes|i\rangle\langle i|\right)\\
    &=\sup_{\Lambda\in\mathrm{CLODCC}}D\left(\sum_i\tr_\mathcal{H}\left(\left(M_\Lambda^i\right)^\Gamma\left(\rho^\Gamma\right)^{\otimes n}\right)\otimes|i\rangle\langle i|||\sum_i\tr_\mathcal{H}\left(\left(M_\Lambda^i\right)^\Gamma\frac{\mathbb{I}^{\otimes n}}{d^{2n}}\right)\otimes|i\rangle\langle i|\right)\\
    &\leq D\left(\left(\rho^\Gamma\right)^{\otimes n}||\left(\frac{\mathbb{I}}{d^2}\right)^{\otimes n}\right)=nD\left(\rho^\Gamma||\frac{\mathbb{I}}{d^2}\right).
    \end{split}
\end{equation}
\end{widetext}
Hence,
\begin{equation}
\begin{split}
    &\lim_{n\rightarrow\infty}\frac{1}{n}\sup_{\Lambda\in\mathrm{CLODCC}}D\left(\Lambda(\rho^{\otimes n})||\Lambda\left(\frac{\mathbb{I}^{\otimes n}}{d^{2n}}\right)\right)\\
    &\leq\lim_{n\rightarrow\infty}D\left(\rho^\Gamma||\frac{\mathbb{I}}{d^2}\right).
    \end{split}
\end{equation}
In the above $\Lambda = \{M_\Lambda\}$ is a POVM of an operation from the set $\mathrm{CLODCC}$. In the second equality we use the identity $\tr (XY) = \tr (X^{\Gamma}Y^{\Gamma})$ for matrices $X$ and $Y$, and the fact that $\left(\rho^{\otimes}\right)^{\Gamma} = \left(\rho^{\Gamma}\right)^{\otimes n}$. The last inequality follows from the fact that the relative entropy is non-increasing under CPTP maps. $\square$

\subsection{Rate of repeated private randomness}
Now we are in position to derive an asymptotic version of the distinguishability bound, that is, the quantity that upper bounds the rate of repeated randomness $\mathcal{R}_A^{\ABC}$. It measures the distinguishability of the state from the maximally mixed state in terms of the relative entropy of the probability distributions that can be obtained by $\mathrm{CLODCC}$. 

We start from presenting a rigorous definition of rates of repeated randomness. Namely, for input states ${\r}_{AC_1}$ between $A$ and $C$, and $\tilde{\rho}_{C_2B}$ between $C$ and $B$, we call
\begin{widetext}
\begin{equation}
\mathcal{R}_A^{A \leftrightarrow C\leftrightarrow B}({\rho}_{AC_1}\ot\tilde{\rho}_{C_2B}):=\inf_{{\epsilon}>0}\limsup_{n\to\infty}\sup_{\Lambda_n \in \mathrm{CLODCC}, \alpha_m}\left\{\frac{m}{n}:\tr_{C}\left(\Lambda_n\left(\left({{\rho}}_{AC_1}\ot\tilde{\rho}_{C_2B}\right)^{\ot n}\right)\right)\approx_{\epsilon}\alpha_{ m}\right\}
\end{equation}
\end{widetext}
the \emph{quantum private randomness repeater rate of $\r$ and $\tilde{\r}$ with respect to arbitrary $\mathrm{CLODCC}$ operations among $A$, $B$ and $C$, that can be obtained on a system $A$}.

Let $\mathrm{CLODCC}(A:B)$ be the set of POVMs which can be implemented with $\mathrm{CLODCC}$ operations. An element of this class is a corresponding CPTP map. That is, instead of a POVM given by $\{M_i\}$, we consider the CPTP map $M: X \mapsto \sum_i (\tr (M_i X))\otimes|i\>\<i|$. Hence, $M(\rho)$ is a distribution of POVMs elements from the stet $\{M_i\}$ measured for a density operator $\rho$. Our bound on the quantum independent randomness repeater rate involves the \textit{measured relative entropy} with respect to the set $\mathrm{CLODCC}$,
\begin{widetext}
\begin{equation}
D_{C \leftrightarrow AB}(\rho_{AC_1}\otimes \tilde\rho_{C_2B}):=\sup_{M \in \mathrm{CLODCC}(C:AB)}D\left( M(\rho\ot \tilde\rho)\|M\left( {\mathbb{I}\over d_{ABC}}\right) \right). 
\label{measured.rel.ent.def}
\end{equation}
\end{widetext}
By $d_{ABC}$ we mean the multiplication of dimensions of $\rho$ and $\tilde\rho$. We denote by $D^\infty_{C \leftrightarrow AB}$ the regularized version of \eqref{measured.rel.ent.def}, analogously to the relationship between \eqref{DS.infty.definition} and \eqref{DS.definition}. 

Before we prove the bound, we need a lemma showing 
 a lower bound on the measured relative entropy distance from the maximally mixed state for states that approximate independent states. We show that the measured relative entropy distance with respect to $\mathrm{U+Deph}$ from  the maximally mixed state is proportional to $m$ on $\rho\approx_\epsilon \alpha_m$.
\begin{lemma} \label{lemma:normalisation} 
    For $\rho\approx_\epsilon \alpha_m^{AA'B}$ of dimensionality $|AA'B|$, we have
    \be
    D_{\mathrm{U+Deph}}(\rho\|{\id \over |AA'B|})\geq (1-\epsilon) m-h(\epsilon).
    \ee
\end{lemma}

{\it Proof}.
We will follow the proof of Lemma from \cite{BCHW} with appropriate changes, since a general idit is twisted coherence rather than entanglement. We use the fact that $\alpha_m$ can be expressed as $U P^{m}_A \otimes \sigma_{A'B'} U^\dagger$. Here $U$ is a controlled unitary operator, with control $A$ and target $A'B'$, while $\sigma_{A'B'}$ is an arbitrary state. Then,
\begin{equation}
\begin{split}
&D_{\mathrm{U+Deph}}(\rho\|{\id \over |AA'B|})\\
&=\sup_{\Lambda\in \{\mathrm{U+Deph}\}}D(\{\tr\left(M_{\Lambda}(\rho)\right)\}\|\{\tr(M_{\Lambda}\left({\id \over |AA'B|}\right)\})\\
&\geq D_{\mathrm{U+Deph}}(\tr_{A'B'}(U\rho U^\dagger) \| \tr_{A'B'}(U{\id \over |AA'B|}U^{\dagger}))\\
&=D_{\mathrm{U+Deph}}(\tilde P^m_A \| {\id \over |A|}) \\
& \geq D_{\mathrm{U+Deph}}(\{\tr(P_{m,F}{\tilde P}^m_A)\} \| \{\tr(P_{m,F} {\id \over |A|})\}) \\
& \geq  (1-\epsilon) m-h(\epsilon),
\end{split}
\end{equation}
where ${\tilde P}^m_A:=\tr_{A'B'}(U\rho U^\dagger)$ is a state, $\epsilon$-close to $P^m_A \equiv \sum_{i,j=0}^{2^m-1} {1\over 2^m}|i\>\<j|_A$.
The first inequality holds due to monotonicity of $D(\cdot||\cdot)$ and the fact that $U \in \{\mathrm{U+Deph}\}$.  The second inequality follows from: (i) monotonicity under the projective measurement $\{P_{m,F}\}$ onto the basis of the Fourier transform of the basis $\{|i\>\}_{i=0}^{2^m-1}$ ($P^A_m$ is an element of this transformed basis), and (ii) $P_{m,F} \in \{\mathrm{U+Deph}\}$. The last inequality is due to $\{\tr (P_{m,F} {\id \over |A|})\} = \{1/2^m\}$. Moreover, $\tr(P_{m,F} {\tilde P}^A_m)\geq 1-\epsilon $, which follows from $\rho \approx_\epsilon \alpha_m$. Further, the highest entropy among distributions 
$\{1-\epsilon,\lambda_1,\ldots,\lambda_{d-1}\}$ is achieved by the most mixed one for $\lambda_i = {\epsilon\over {d-1}}$. We thus obtain the lower bound on the relative entropy of the distribution, as claimed. \qed

\medskip
We now come to the main result of this Section.
\begin{theorem} \label{theorem:fundamental}
For all states $\rho_{AC_1}$ and $\tilde\rho_{C_2B}$:
    \begin{equation}
    \mathcal{R}_A^\ABC(\rho_{AC_1}\otimes \tilde\rho_{C_2B})\leq D_\CAB^\infty(\rho_{AC_1}\otimes \tilde\rho_{C_2B}). 
    \end{equation}
\end{theorem}
{\it Proof}. For any $\epsilon >0$, by the definition of the rate of repeated private randomness, there exists $n\in\mathbb{N}$ and $\Lambda \in \beta:=\mathrm{CLODCC}(A^n : C^n : B^n)$, such that $r \geq \mathcal{R}_A^{\ABC}(\rho_{AC_1}\otimes \tilde\rho_{C_2B})-\epsilon$ and 
$\tilde\alpha:=\tr_C\Lambda((\rho_{AC_1}\otimes \tilde\rho_{C_BB})^{\otimes n})\approx_\epsilon \alpha_{\lfloor nr \rfloor} $, where $\lfloor \cdot \rfloor$ denotes the floor function. Taking $\sigma_{ABC} = {\mathbb{I}\over |{ABC}|}$ and $\tilde\sigma:=\tr_C\Lambda(\sigma )$, we have
\begin{equation} 
\begin{split}
    &\max_{M \in \beta}D(M(\rho_{AC_1}^{\otimes n}\otimes \tilde\rho_{C_2B}^{\otimes n})\|M(\sigma_{ACB}))\\
&\geq   \max_{M \in \beta}D(M( \Lambda(\rho_{AC_1}^{\otimes n}\otimes \tilde\rho_{C_2B}^{\otimes n}))\|M(\Lambda(\sigma_{ACB} ))) \\
& \geq \max_{M \in \mathrm{U+Deph}}D(M(\tilde\alpha_{AB})\|M(\tilde\sigma_{AB})).
\end{split}
\label{b1}
\end{equation}
Thanks to Observation \ref{obs:CLODCCinclusions} and assumption that $\Lambda \in \mathrm{CLODCC}(C:AB)$ we obtain the first inequality. 
The third line follows from the fact that we restrict maximisation to the set of operations that are induced on system $AB$ from a $\mathrm{CLODCC}(C:AB)$ via trace over $C$. The set of these operations is denoted by $S$.  Due to Observation \ref{obs:induced}, the set $S$ includes $\mathrm{U+Deph}$. We  get the lower quantity if we restrict supremum to the operations from $\mathrm{U+Deph} \subsetneq S$. Due to Observation~\ref{obs1}, $\tilde\sigma = {\mathbb{I}\over d_{out}}$,
where $d_{out}$ is the dimension of the output of the map $\tr_C\Lambda(\cdot)$. 

Applying Lemma \ref{lemma:normalisation} with $\tilde\alpha\approx_\epsilon \alpha_{\lfloor nr\rfloor}$ we arrive at
\be 
\label{b2}
\max_{M \in\mathrm{U+Deph}}D(M(\tilde\alpha_{AB})\|M(\tilde\sigma_{AB}))\geq (1-\epsilon) \lfloor nr \rfloor-h(\epsilon).
\ee 
Bounds~\eqref{b1} and~\eqref{b2}, together with minimization over $\sigma$ and taking the limit $n\rightarrow \infty$, imply the following lower bound on $D^{\infty}_\CAB$:
\be 
D_\CAB^\infty(\rho_{AC_1}\otimes \tilde\rho_{C_2B})\geq (1-\epsilon) r.
\ee
Taking into account that $r \geq \mathcal{R}_A^{\ABC}(\rho_{AC_1}\otimes \tilde\rho_{C_2B})-\epsilon$ with arbitrary $\epsilon$, the statement is proved. \qed 

{\cor \label{cor:boundsRelEnt}
    The following inequality holds for all PPT states $\rho=\rho_{C_1A}$ and $\tilde\rho=\tilde{\rho}_{C_2B}$:
    \be
    \label{eq:cor:boundsRelEnt}
    \mathcal{R}_A^{\ABC}(\rho \otimes \tilde{\rho})\leq D\left( \rho^\Gamma || {\1\over |AC_1|}\right) +D\left( \tilde{\rho}^\Gamma ||{\1\over |C_2B|}\right),
    \ee
    where $ d_{\rho},d_{\tilde\rho}$ stand for the dimensions of $\rho,\tilde\rho$ respectively.
}

This Corollary follows from applying Theorems~\ref{eq:dist-thm} and \ref{theorem:fundamental} to $J=\mathrm{CLODCC}(C:AB)$.

From the bound~\eqref{eq:cor:boundsRelEnt} in Corollary~\ref{cor:boundsRelEnt} we can conclude that there are states that have localisable randomness equal almost 1, while their repeated independent randomness is vanishingly small (see Section~\ref{sec:examples} for examples).

To interpret the above result, we should compare the localisable and repeated private randomness. Theorem \ref{thm:neg-coh-random} of \cite{YHW}, invoked in Section \ref{preliminaries}, states that localisable private 
randomness of an input state $\rho_{AC_1}$ is equal to its global purity, i.e., $\log |AC_1| - S(AC_1)_\rho$. Using the equality of $\log |AC_1|$ for $\rho_{AC_1}$ and for $\rho_{AC_1}^{\Gamma}$, the RHS of (\ref{eq:cor:boundsRelEnt}) can be rewritten as $\log|AC_1|-S(AC_1)_{\rho^\Gamma} + \log|C_2B|-S(C_2B)_{\tilde{\rho}^\Gamma}$. However, for any state $\sigma_{AC_1}$, $\log |AC_1| - S(AC_1)_\sigma = (\log |A| - S(A)_\sigma) + (\log|C_1| - S(C_1)_\sigma) + I(A:C_1)_\sigma$. That is, the global purity can be split into purity accessible locally (sum of the first two terms), and the correlation part (the mutual information).
The locally accessible purity is a type of private randomness that is accessible to $A$ and $B$ {\it without} help of $C$, and hence is always available in our private randomness repeater scenario. The partial transposition does not change entropy of local subsystem, 
$S(A)_\rho = S(A)_{\rho^{\Gamma}}$, and the same holds for $B$. Hence, for $\tilde{\rho} = \rho$, the difference between localisable private randomness from $\rho$ at system $A$ and our bound reads:
\begin{widetext}
\begin{equation}
\begin{split}
    &\log |AC_1| - S(AC_1)_\rho - (\log |AC_1| - S(AC_1)_{\rho^\Gamma}) -(\log |C_2B| - S(C_2B)_{\tilde{\rho}^\Gamma}) \\ 
    &=I(A:C_1)_\rho - (\log|B| -S(B)_{\tilde \rho} + \log|C_2| -S(C_2)_{\tilde \rho} + I(A:C_1)_{\rho^\Gamma} + I(C_2:B)_{\tilde{\rho}^{\Gamma}}).
    \end{split}
		\label{equation.bound.logSlog}
\end{equation}
\end{widetext}
Thus, due to the term $\log |B| -S(B)_\rho + \log|C_2| -S(C_2)_\rho$ appearing on the RHS of \eqref{equation.bound.logSlog}, the above bound is weak for states that contain local purity. However, as we will see, it is sufficiently powerful for all states that have local purity equal to zero, i.e., that have both subsystems in maximally mixed states. In the latter case, considering also $\rho_{AC_1}={\tilde \rho}_{C_2B}$, the gap between localisable and repeated localisable randomness reads $I(A:B)_\rho - 2I(A:B)_{\rho^\Gamma}$. In  Section~\ref{sec:examples} we will study behaviour of this gap for the family of separable Werner states. 

\section{Direct bound for PPT states is not tighter than the indirect one}
\label{sec:direct}
In this Section we provide a more direct proof of Corollary \ref{cor:boundsRelEnt}. One might think that the latter bound could be improved by getting rid of the factor $2$ in front of the one presented in~\eqref{eq:corollary} in Section \ref{sec:summary},
 as analogous phenomenon happens for the private key (see Lemma $12$ and Theorem $13$ of the Supplemental Material of \cite{BCHW}). As we will see below, this is not the case: we obtain the same bound. We show it here, because its intermediate step is worth mentioning separately. It states that the repeated private randomness is upper bounded for states from PPT set by its value taken on the partially transposed state:

\begin{theorem} For any two bipartite states $\rho$ and ${\tilde \rho}$ that have
positive partial transposition,
\begin{equation}
{\cal R}_A^{A\leftrightarrow C_1C_2 \leftrightarrow B}(\rho_{AC_1}\otimes {\tilde \rho}_{C_2B}) \leq {\cal R}_A^{A\leftrightarrow C_1C_2 \leftrightarrow B}(\rho_{AC_1}^{\Gamma}\otimes {\tilde \rho}_{C_2B}^{\Gamma}).
\end{equation}
    \label{thm:simple-bound}
\end{theorem}

{\it Proof}.
 We first note that the definition of ${\cal R}_A^{A\leftrightarrow C\leftrightarrow B}$ involves the term
$\tr_C \Lambda((\rho_{AC_1}\otimes \rho_{C_2B})^{\otimes n})$, with $\Lambda \in \mathrm{CLODCC}(A:C:B) \subset \mathrm{LOCC}(A:C:B) \subset \mathrm{SEP}(A:B:C)$, where $\mathrm{SEP}(A:B:C)$ are the operations that can be expressed in a form $\sum_i A_i \otimes B_i \otimes C_i (\cdot)A_i^{\dagger} \otimes B_i^{\dagger} \otimes C_i^{\dagger}$.
Adopting the idea of the proof of Lemma $12$ from \cite{BCHW}, we note that $\tr_C(\sigma_{ACB})=\tr_C((\id_{AB}\otimes T_C) \sigma_{ABC})$, i.e., we can transpose the state on system $C$ before tracing it, then trace and obtain the original state traced over system $C$. This fact holds for any state $\sigma$, and in particular for $\sigma := \Lambda(\rho_{AC_1}\otimes {\tilde{\rho}}_{C_2B})$. Hence,
\begin{widetext}
\begin{equation}
\begin{split}
\tr_C\Lambda((\rho_{AC_1}\otimes \rho_{C_2B})^{\otimes n})&=\tr_C ((\id_{AB} \otimes T_C)\Lambda((\rho_{AC_1}\otimes \rho_{C_2B})^{\otimes n}))\\
&=\tr_C ((\id_{AB} \otimes T_C)\sum_{ijk}A_i\otimes B_j \otimes C_k (\rho_{AC_1}\otimes \rho_{C_2B})^{\otimes n} A_i^{\dagger}\otimes B_j^{\dagger} \otimes C_k^{\dagger}).
\end{split}
\end{equation}
\end{widetext}
Using $(\id \otimes T)(X_1 \otimes X_2 \rho Y_1 \otimes Y_2)=X_1 \otimes Y_2^T (\rho^{\Gamma}) Y_2 \otimes X_2^T$, we obtain:
\begin{widetext}
\begin{equation}
\begin{split}
    &\tr_C ((\id_{AB} \otimes T_C)\sum_{ijk}A_i\otimes B_j \otimes C_k (\rho_{AC_1}\otimes \rho_{C_2B})^{\otimes n} A_i^{\dagger}\otimes B_j^{\dagger} \otimes C_k^{\dagger})\\
    &=\tr_C (\sum_{ijk}A_i\otimes B_j \otimes C_k^* (\rho_{AC_1}^{\Gamma}\otimes \rho_{C_2B}^{\Gamma})^{\otimes n} A_i^{\dagger}\otimes B_j^{\dagger} \otimes (C_k^*)^{\dagger}).
    \end{split}
\end{equation}
\end{widetext}
We will show now, that $C_k^*$ are such, that the total operation
$\sum_{ijk} A_i\otimes B_j \otimes C_k^*(\cdot)A_i^{\dagger}\otimes B_j^{\dagger} \otimes (C_k^*)^{\dagger}$ is a valid $\mathrm{CLODCC}(C:AB)$ operation.  Let $\hookrightarrow |0\>\<0|_X$ denote the operation of adding an ancillary state $|0\>$ to the system $X$.
Any operation from $\mathrm{CLODCC}(C:AB)$ can be simulated by the following four LOCC operations (and their composition in a proper order):
\begin{enumerate}
    \item Unitary transformation on system $C$: $U_{C\rightarrow C'c}$.
    \item Dephasing channel from $C$ to $A$,
    i.e. $\{P^i_c\otimes \id_{C'}\}_{i=0}^{|c|-1}$ with $P^i:= |i\>\<i|_c$.
        \item Operation which changes the system $c$ in a way that it is in the same state as some dephased system on $AB$. It first adds an ancillary blank state, and further performs appropriate shift  $\{S^c_{i,|c|}\}_{i=0}^{|c|-1}\circ \hookrightarrow |0\>\<0|_c$, with
        $S^c_{i,|c|}|j\> := |j+i\, \mathrm{mod} |c|\,\>$. This operation is controlled by the outcomes of $\{P^i_{a}\otimes \id_{\bar a}\}_{i=0}^{|a|-1}$ with $P^i_a:= |i\>\<i|_a$, $a$ being an arbitrary subsystem of $AB$ satisfying $|a|=|c|$, and ${\bar a}$ denoting complement of $AB$ to $a$.
        \item $\tr_c$ (used only after a dephasing channel and an operation on system $A$ analogous to the 3rd operation on this list).
\end{enumerate}
For any $k$ there is $C_k= M_1\circ \cdots \circ M_l \circ \cdots$, where $M_l$ are Kraus' operators from the above set of operations (up to restriction that $\tr_c$ can be used only after $3$rd operation from the list). Hence, $C_k^*= M_1^*\circ \cdots \circ M_l^* \circ \cdots$.
All operations on the above list, apart from the $1$st, do not change under complex conjugation, as they are formulated with real numbers, while $U_{C\rightarrow C'c}$ becomes another unitary transformation $U^*_{C\rightarrow C'c}$. Thus, any $\mathrm{CLODCC}(C:AB)$ operation $\Lambda$ after partial transposition $(\cdot)^\top_C\otimes \id_{AB}$ becomes some other operation $\Lambda' \in \mathrm{CLODCC}(C:AB)$. By evaluating it on $\rho_{AC_1}^{\Gamma}\otimes\rho_{C_2B}^{\Gamma}$, the assertion follows. $\square$

\begin{remark}
Although the fact that $\mathrm{CLODCC}(A:B)\subsetneq \mathrm{LOCC}(A:B)$ was already noticed in the context of resource theory of purity \cite{huge-delta},
the above simulation of an operation from $\mathrm{CLODCC}$ by means of 
$\mathrm{LOCC}$ is an explicit proof of this inclusion. Local operations
of enlarging system $\hookrightarrow |0\>\<0|$, partial trace, and von Neumann projection, are explicitly inside $\mathrm{LOCC}$. The operation of application of 
the shift $S_{i,|s|}$ is controlled by the outcome of the projective measurement
on the other system, which employs the communication based inter-dependencies of the Kraus operators of an $\mathrm{LOCC}$ operation.
\end{remark}

From the above we have an immediate Corollary, where
by $\mathrm{G}(\rho_{AB})$ we denote $\log|AB|- S(AB)_\rho$. 

\begin{cor} For any two bipartite states $\rho$ and ${\tilde \rho}$ that have positive partial transposition, there is:
\begin{equation}
{\cal R}_A^{A\leftrightarrow C_1C_2 \leftrightarrow B}(\rho_{AC_1}\otimes {\tilde \rho}_{C_2B}) \leq  \mathrm{G}(\rho_{AC_1}^{\Gamma}\otimes{\tilde\rho}_{C_2B}^{\Gamma}).
\end{equation}
\label{cor:simple-bound}
\end{cor}
{\it Proof}.
We first note that $\mathrm{CLODCC}(A:C_1C_2:B)\subset \mathrm{CLODCC}(A:(C_1C_2B))$  (see Observation \ref{obs:CLODCCinclusions}). The state $\sigma= \rho_{AC_1}^{\Gamma}\otimes\rho_{C_2B}^{\Gamma}$, treated as a bipartite state with a partition $A:(C_2C_1B)$, has a positive partial transposition, since $\rho_{AC_1}$ has it positive by assumption. Hence Theorem \ref{thm:neg-coh-random} implies that $G(\sigma)$ is achieved. $\square$

Since $G(\rho \otimes {\tilde{\rho}})$ is additive on tensor product, the RHS of \eqref{cor:simple-bound} is equal to the RHS of the bound \eqref{eq:cor:boundsRelEnt} of Corollary \ref{cor:boundsRelEnt}. So, the above bound is no better than already presented one. This is in contrast with the case of private key \cite{BCHW}, where the corresponding bound was better by factor of $2$ (c.f. Lemma $12$ and Theorem $13$ of the Supplemental Material of \cite{BCHW}).

\section{Limitation for i.i.d. private randomness repeaters for some ibits}
\label{sec:limitations_iid}
In this Section we focus on a simpler case in which the three parties first perform the same $\mathrm{CLODCC}$ operation on each of the copies of the state, and then $A$ and $B$ perform general $\mathrm{CLODCC}(A:B)$.
We begin with defining the rate of repeated private randomness gained by $\mathrm{CLODCC}(C^{\mbox{iid}}:(A^{\mbox{iid}}:B^{\mbox{iid}}))$ operations. As we will see, in this case even some states with negative partial transposition will have limited repeated private randomness.

We begin with a formal definition of private randomness repeater based on the operations mentioned above.
\begin{widetext}
\begin{align}
\begin{split}
\mathcal{R}_A^{C^{\mbox{iid}}:(A^{\mbox{iid}}:B^{\mbox{iid}})}({\rho}_{AC_1}\ot\tilde{\rho}_{C_2B}):=\inf_{\substack{{\epsilon}>0}}\limsup_{n\to\infty}\sup_{\Lambda_n \in \mathrm{CLODCC}(C^{\mbox{iid}}:(A^{\mbox{iid}}:B^{\mbox{iid}})), \alpha_m}\left\{\frac{m}{n}:\tr_{C}\Lambda_n\left(\left({{\rho}}_{AC_1}\ot\tilde{\rho}_{C_2B}\right)^{\ot n}\right)\approx_{\epsilon}\alpha_{ m}\right\}
\end{split}
\end{align}
\end{widetext}
will be called the \emph{quantum i.i.d. private randomness repeater rate of $\r$ and $\tilde{\r}$ with respect to $\mathrm{CLODCC}(C^{\operatorname{iid}}:(A^{\operatorname{iid}}:B^{\operatorname{iid}}))$ operations among $A$, $B$, and $C$, that can be obtained at system $A$}. With a little abuse of notation we will denote 
$\mathcal{R}_{C^{\operatorname{iid}}:(A^{\operatorname{iid}}:B^{\operatorname{iid}})}$
as $\mathcal{R}_A^{\operatorname{iid}}$. Moreover, in case of $\rho=\tilde{\rho}$, we will refer to 
$\mathcal{R}_A^{\operatorname{iid}}(\rho\otimes \tilde{\rho})$ as to
$\mathcal{R}_A^{\operatorname{iid}}(\rho)$.

From Lemma $1$ of the content of the Supplementary Note 2 in \cite{BCHW},
we know 
{\cor For any two states $\rho_{AC_1}$ and $\tilde{\rho}_{C_2B}$ and any $\Lambda \in \mathrm{CLODCC}(A:C_1 C_2:B)$, the output state $\hat{\rho}_{AB}=\tr_C\Lambda(\rho_{AC_1}\otimes \rho_{C_2B})$, satisfies
\begin{equation}
    ||\hat{\rho}_{AB} -{\id \over 
    |AB|_{\hat{\rho}}}||_1 \leq 
    ||\rho_{AC_1}^{\Gamma} - {\id\over |{AC_1}|}||_1 +    ||\tilde{\rho}_{C_2B}^{\Gamma} - {\id\over |{C_2B}|}||_1. 
\end{equation}
\label{cor:norms}
} 
\begin{prooftw} 
{\it Proof.} Follows from ${\id\over d }\in SEP$ and
$\mathrm{CLODCC} \subset \mathrm{LOCC}$, as a special case
of Lemma $1$ in \cite{BCHW}. \qed
\end{prooftw}

{\proposition For a state  $\rho \in \mathbb{C}^d\otimes \mathbb{C}^d$, satisfying $||\rho^{\Gamma} -{\id \over d}||_1\leq {1\over \mathrm{e}}$, and any operation $\Lambda \in \mathrm{CLODCC}(A:C_1C_2:B)$, the output state $\hat{\rho}_{AB} = \tr_C \Lambda(\rho\otimes \rho)$ satisfies
\begin{equation}
    |\log|AB|_{\hat{\rho}} - S(AB)_{\hat{\rho}}| \leq  2||\rho^{\Gamma} - {\id\over d}||_1\log d + \eta(2||\rho^{\Gamma}-{\id\over d}||_1),
\label{prop.two.ineq}
\end{equation}
where $\eta(x):=-x \log x$.
\label{prop:entropy-bounded}
} 
\begin{prooftw}
{\it Proof}. From the asymptotic continuity of quantum mutual information \cite{Alicki_2004,Shirokov_2017} for any $\rho,\rho' \in \mathbb{C}^{d_A}\otimes \mathbb{C}^{d_B}$ such that $||\rho - \rho'||_1 \leq \epsilon$ with $0<\epsilon <{1\over \mathrm{e}} \approx 0.368$, one has
\begin{equation}
    |S(AB)_{\rho} - S(AB)_{\rho'}|\leq \epsilon \log d_{AB} + \eta(\epsilon).
\end{equation}
Since the von Neumann entropy of the maximally mixed state equals $\log |AB|$, the assertion follows directly from Corollary \ref{cor:norms}. \qed
\end{prooftw}

We will exemplify the upper bound \eqref{prop.two.ineq} using the independent state from~\eqref{eq:alpha1}.  This state has negative partial transposition.
We will use the property
\begin{equation}
    ||\alpha_{V,d}^{\Gamma} - {\id\over 2d^2}||_1 \leq {2\over d}.
    \label{eq:swap-bound}
\end{equation}
We are ready to show the gap between private randomness and repeated private randomness for $\alpha_{V,d}$, for sufficiently large $d$.

\begin{theorem}
    The family of states $\{\alpha_{V,d}\}_{d=2}^{\infty}$ satisfies the following properties:
    \begin{enumerate}
        \item For $d > 2$, $R_A(\alpha_{V,d})= R_B(\alpha_{V,d})= 1$.
        \item For $d>11$, 
        \begin{equation}
            \mathcal{R}_A^{\operatorname{iid}}(\alpha_{V,d})\leq {{4\log d}\over d} + \eta({{4}\over d}).
        \end{equation}
        \item For $d > 32$, $1=R_A(\alpha_{V,d}) = R_B(\alpha_{V,d}) > \mathcal{R}_A^{\operatorname{iid}}(\alpha_{V,d})\rightarrow_{d\rightarrow \infty} 0.$
    \label{item:last}
    \end{enumerate}
    \label{thm:iid-bound}
\end{theorem}
\begin{prooftw}
{\it Proof}. The first statement follows from negativity of coherent information of $\alpha_{V,d}$ for $d>2$, so that Theorem \ref{thm:neg-coh-random} applies.
Let us denote $\alpha_{V,d}$ as $\alpha_{AA'B'}$ to indicate subsystems explicitly. Like for the states with positive partial transposition, the conditional entropy
$S(B'|AA')$ equals to the {\it global purity} of $\alpha_{AA'B'}$, i.e., to $\log |AA'B'| - S(AA'B')_{\alpha_{V,d}}$. This in turn gives $I(AA':B')_{\alpha_{V,d}} =1$.

For the second statement, we focus on a perspective of party $A$.
This property follows from the sequence of inequalities:
\begin{equation}
\begin{split}
    \mathcal{R}_A^{\mbox{iid}}(\alpha\otimes\alpha)&\leq R_A(\hat{\rho}) \leq \log|AB|_{\hat{\rho}} - S(AB)_{\hat{\rho}}\\ &\leq 2||\rho^{\Gamma} - {\id\over d}||_1\log d + \eta(2||\rho^{\Gamma}-{\id\over d}||_1)\\
    &\leq {{4\log d}\over d} + \eta({{4}\over d}), 
    \end{split}
\end{equation}
where $\hat{\rho} = \tr_C\Lambda(\alpha_{AC_1}\otimes\alpha_{C_2B})$ with $\Lambda \in \mathrm{CLODCC}(A:C_1C_2:B)$.
The first inequality comes from the definition of the class of operations involved in $\mathcal{R}_A^{\operatorname{iid}}$. Second one holds because private randomness cannot be greater than the global purity, i.e., the amount of purity that $A$ and $B$ can obtain when they join their systems and act globally. The value of global purity
is achievable due to the Schumacher compression \cite{Schumacher1995,Nielsen-Chuang}. The next inequality follows from Corollary \ref{cor:norms}. The last one is due to Eq. (\ref{eq:swap-bound}) and the fact that for $d>11$, we have $2\times {2\over d} \leq {1\over \mathrm{e}}$ and the Proposition \ref{prop:entropy-bounded}. For $d > 32$, the RHS of
just proven bound is less than $1$, i.e., less than $R_A(\alpha_{V,d})$. The argument for $R_B$ is symmetric. \qed
\end{prooftw}

 The inequality presented in the \ref{item:last}rd item of Theorem \ref{thm:iid-bound}, seems to be trivial, as $R_A$ involves in its definition a class of operations not restricted by ``i.i.d.''. However, we can make sure that this is not the case for the states $\alpha_{V,d}$ on systems $AA'$. Indeed, for these states, private randomness is directly accessible for Alice via identical measurements on each copy of $\alpha_{V,d}$ on subsystem $A$. One can then define $R_A^{\operatorname{iid}}$ as private randomness localisable at subsystem of party $A$ via identical operations on the input state.

\begin{cor}
For system $A$ of the state $\alpha_{V,d}$
with $d > 32$, there is 
\begin{equation}
    \mathcal{R}_A^{\operatorname{iid}}(\alpha_{V,d}) < R_A^{\operatorname{iid}}(\alpha_{V,d}).
\end{equation}
\end{cor}

\section{A gap between localisable and repeated private randomness for separable Werner states}
\label{sec:examples}
In this Section we show that the main result holds for a larger set
of Werner states than the fully symmetric state and we briefly study
the critical dimension for which there is a limitation in the randomness
repeaters.

A general Werner state $\rho$ is a convex combination
\begin{equation}
\rho=\left(1-\theta\right)\rho_{s}+\theta\rho_{a},
\label{eq:wernerstate}
\end{equation}
with $\theta$ the mixing parameter, the symmetric state $\rho_{s}:=\frac{1}{d^{2}+d}\left(\id+V\right)$, the antisymmetric state $\rho_{a}:=\frac{1}{d^{2}-d}\left(\id-V\right)$, where $V$ is the swap operator, while $d$ is the dimension of the systems $A$ and $B$. We will be using the facts that the partial
transpose $\left(\cdot\right)^{\Gamma}$ is a linear operator, $\left(\id\right)^{\Gamma}=\id$, and $\left(V\right)^{\Gamma}=d\left|\Phi^{+}\right\rangle \left\langle \Phi^{+}\right|$.
Using the above results, we write the partial transposes of $\rho_{s}$ and $\rho_{a}$ as $\left(\rho_{s}\right)^{\Gamma}=\frac{\id-\left|\Phi^{+}\right\rangle \left\langle \Phi^{+}\right|}{d^{2}+d}+\frac{\left|\Phi^{+}\right\rangle \left\langle \Phi^{+}\right|}{d}$, $\left(\rho_{a}\right)^{\Gamma}=\frac{\id-\left|\Phi^{+}\right\rangle \left\langle \Phi^{+}\right|}{d^{2}-d}-\frac{\left|\Phi^{+}\right\rangle \left\langle \Phi^{+}\right|}{d}$.
Defining $\left|\Phi^{+}\right\rangle \left\langle \Phi^{+}\right|^{\bot}:=\id-\left|\Phi^{+}\right\rangle \left\langle \Phi^{+}\right|$,
we get $\rho^{\Gamma}=\left(1-2\theta\right)\frac{\left|\Phi^{+}\right\rangle \left\langle \Phi^{+}\right|}{d}+\left[\frac{\theta}{d^{2}-d}+\frac{1-\theta}{d^{2}+d}\right]\left|\Phi^{+}\right\rangle \left\langle \Phi^{+}\right|^{\bot}\label. $.
The state $\rho^{\Gamma}$ is  diagonal in basis of maximally entangled states, called a {\it Bell basis} \cite{werner-alltel} because it is a convex combination
of Bell diagonal states $\rho_{s}^{\Gamma}$ and $\rho_{a}^{\Gamma}$. From
the form of  $\rho^{\Gamma}$ one can directly obtain the eigenvalues of $\rho^{\Gamma}$
in the Bell basis:
\begin{equation}
\begin{split}
&\lambda_{0}= \frac{\left(1-2\theta\right)}{d},\\
&\lambda_{1}=\ldots=\lambda_{d^{2}-1}= \frac{1}{d}\left[\frac{\theta}{d-1}+\frac{1-\theta}{d+1}\right].
\end{split}
\label{eq:eigens}
\end{equation}
The eigenvalue $\lambda_{0}$ is associated with the eigenvector $\left|\Phi^{+}\right\rangle $,
while all other $d^{2}-1$ eigenvalues are equal and given by (\ref{eq:eigens}).
Because $\rho^{\Gamma}$ is Bell diagonal, the reduction to individual systems $A$ and $B$ gives the maximally mixed state, hence $S\left(A\right)_{\rho^{\Gamma}}=S\left(B\right)_{\rho^{\Gamma}}= \log d $.
Computing the entropy of the whole state, which is $S\left(AB\right)_{\rho^{\Gamma}}=\frac{\alpha}{d}\log\left[\frac{1}{\alpha}\left(\frac{d-\alpha}{d^{2}-1}\right)\right]-\log\left[\frac{d-\alpha}{d^{2}-1}\right]+\log d$, where  $\alpha\equiv 1-2\theta$, we are in the position to compute the mutual information:
\begin{eqnarray*}
I\left(A:B\right)_{\rho^{\Gamma}} =\log\left[\frac{d\left(d-\alpha\right)}{d^{2}-1}\right]+\frac{\alpha}{d}\log\left[\frac{\alpha\left(d^{2}-1\right)}{d-\alpha}\right].
\end{eqnarray*}
In consequence,
\begin{equation}
\lim_{d\rightarrow+\infty}I\left(A:B\right)_{\rho^{\Gamma}}=0.
\end{equation}

As noticed in Section I.A, the states 
illustrating our claim are those which satisfy 
\begin{equation}
I\left(A:B\right)_{\rho}>2I\left(A:B\right)_{\rho^{\Gamma}}.\label{eq:condition}
\end{equation}

Let us notice that states $\rho_s$ and $\rho_a$ have supports, respectively, in the orthonormal subspaces $\mathcal{H}_s$ and $\mathcal{H}_a$ of the full Hilbert space $\mathcal{H}_{AB}=\mathcal{H}_s\oplus \mathcal{H}_a$. The von Neumann  entropy of the density matrix $\rho=(\frac{1+\alpha}{2})\rho_s+(\frac{1-\alpha}{2})\rho_a$ reads (see Eq. (12.19) in~\cite{BengtssonZyczkowski-book}):
\begin{equation}
\begin{split}
S(AB)_{\rho}&=h\left(\frac{1-\alpha}{2}\right)+\frac{1+\alpha}{2}S(AB)_{\rho_s}+\frac{1-\alpha}{2}S(AB)_{\rho_a}\\
&=h\left(\frac{1-\alpha}{2}\right)+\frac{1+\alpha}{2}\log(d_+)+\frac{1-\alpha}{2}\log(d_-),
\end{split}
\end{equation}
where $d_+=d(d+1)/2$ and $d_-=d(d-1)/2$.
Hence, the mutual information of the Werner state $\rho$ in the form (\ref{eq:wernerstate}) is
\begin{equation}
I\left(A:B\right)_{\rho}=\log\left[\frac{2d}{\left(d-1\right)^{\left(\frac{1-\alpha}{2}\right)}\left(d+1\right)^{\left(\frac{1+\alpha}{2}\right)}}\right]-h\left(\frac{1-\alpha}{2}\right).
\end{equation}
Hence, $\lim_{d\rightarrow+\infty}I\left(A:B\right)_{\rho}=1-h\left(\frac{1-\alpha}{2}\right)$. This shows that there always exists a value of $d$ large enough to satisfy the condition
(\ref{eq:condition}). The minimum value of $d$ for which the Werner
state $\rho$ satisfies (\ref{eq:condition}) will be called the \emph{critical dimension}
$d_{cri}$. 

\begin{figure}
{\vskip0.4cm}\protect\includegraphics[scale=0.67]{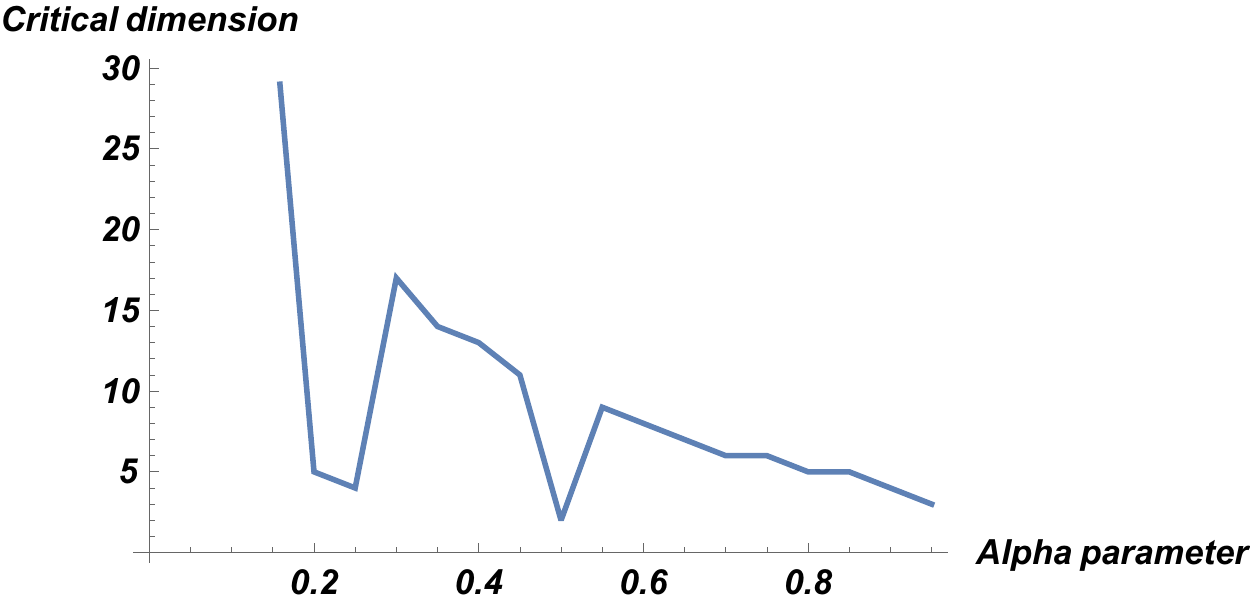}
\caption{\label{fig:three}
The values of $d_{cri}$ as the parameter $\alpha$
increases from $0.1$ by steps of $0.05$. It is worth to mention
some special values, such as $\alpha=0.1$, for which the $d_{cri}$
takes very large value of $51$, and $\alpha\in\{ 0.2,0.5\}$
that determine sudden drops of $d_{cri}$ to the values $\left\{ 5,2\right\}$, respectively.}
\end{figure}

To understand better the nonlinear dependence of $d_{cri}$, we also investigate the plot of $I\left(A:B\right)_{\rho}$
and $2I\left(A:B\right)_{\rho^{\Gamma}}$ versus dimension for some
selected values of $\alpha$, as shown in Figure \ref{fig:three}. The inspection of the sequence presented in Figure \ref{fig:three} shows that the parameter $\alpha$ essentially induces compression of both curves towards the $y$ axis, which generates the different crossing of the curves as $\alpha$ approaches 1. For the values $\alpha$ greater than $0.5$ the value of $d_{cri}$ goes down smoothly and without sudden drops and rises. 

\begin{widetext}

\begin{figure}
\protect\includegraphics[scale=0.5]{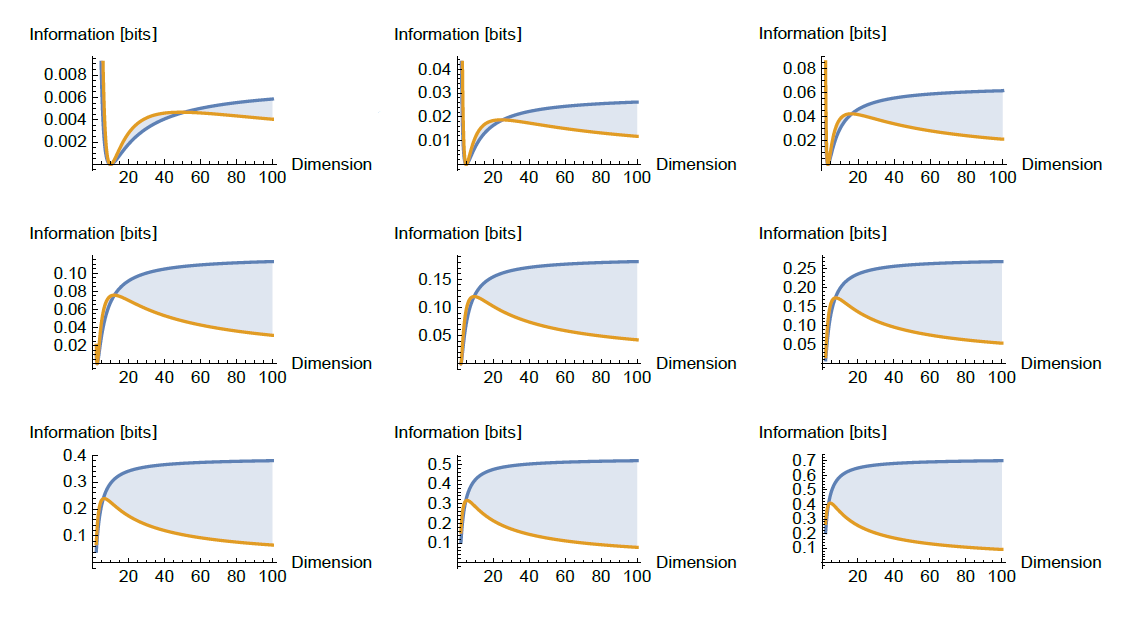}
\caption{
The plots of information vs dimension show the values of $I\left(A:B\right)_{\rho}$
(blue line) and $2I\left(A:B\right)_{\rho^{\Gamma}}$ (orange line)
for several values of $\alpha$, starting with $\alpha=0.1$ in
the upper left corner and increasing by steps of $0.1$ until $\alpha=0.9$
in the bottom right corner. The solid area highlights the gap between
$I\left(A:B\right)_{\rho}$ and $2I\left(A:B\right)_{\rho^{\Gamma}}$.}
\end{figure}
\end{widetext}

\section{towards 2-qubit examples}
\label{2qubitsexamples}
So far the exemplary states were
of dimension higher than $2\otimes 2$.
In this section we show that a wide class of a well
known family of states, that of {\it Bell diagonal } states, (after partial transposition) escapes our technique.

Any Bell diagonal state can be expressed in form of a matrix:
\be
\rho_{Bell} = {1\over 2}\left[\bea{cccc}
a_+ + a_- &0 &0 & a_+ - a_- \\
0 &b_+ + b_- &b_+ - b_- & 0 \\
0 &b_+ - b_- &b_+ + b_- & 0 \\
a_+ - a_- &0 &0 & a_+ + a_- \\
\eea
\right],
\label{eq:Bell-states}
\ee
where the entries are weights of appropriate Bell states:
$\rho_{\mathrm{Bell}}= a_+|\psi_+\>\<\psi_+|+a_-|\psi_-\>\<\psi_-|+ b_+|\phi_+\>\<\phi_+|+ b_-|\phi_-\>\<\phi_-|$.
After partial transposition we obtain
desired family of states:
\be
\rho_{Bell}^{\Gamma} = {1\over 2}\left[\bea{cccc}
a_+ + a_- &0 &0 &b_+ - b_-  \\
0 &b_+ + b_- & a_+ - a_-& 0 \\
0 &a_+ - a_- &b_+ + b_- & 0 \\
b_+ - b_- &0 &0 & a_+ + a_- \\
\eea
\right],
\label{eq:Bell-states}
\ee
Since $\Gamma$ is an involution,
we have $(\rho_{\mathrm{Bell}}^{\Gamma})^\Gamma=\rho_{\mathrm{Bell}}$. We check now, if
some of the states of the form 
$\rho_{\mathrm{Bell}}^{\Gamma}\equiv \rho_{\mathrm{BellG}}$ exhibit
a gap between localisable and repeated randomness. Note, that every Bell diagonal state has maximally mixed subsystems. Since partial transposition does not change the entropy of the subsystems the same holds for $\rho_{\mathrm{BellG}}$. Hence, the condition $I(A:B)_{\rho_{\mathrm{BellG}}} > 2I(A:B)_{\rho_{\mathrm{BellG}}^{\Gamma}}$ is equivalent to: $S(AB)_{\rho_{\mathrm{BellG}}}<2S(\rho_{\mathrm{BellG}}^{\Gamma})-2$, i.e. 
\begin{equation}
    S(AB)_{\rho_{\mathrm{BellG}}} < 
    2S(AB)_{\rho_{\mathrm{Bell}}} - 2
    \label{eq:gap}
\end{equation}
This condition is equivalent to the 
following one:
\begin{eqnarray}
    2 H(\{a_+,a_-,b_+,b_-\}) -2 >\nonumber\\
    H(\{{1\over 2}\left(a_++a_-+b_+-b_-\right),{1\over 2}\left(a_++a_--b_++b_-\right)\nonumber\\
    {1\over 2}\left(a_+-a_-+b_++b_-\right),{1\over 2}\left(-a_++a_-+b_++b_-\right)\})
    \label{eq:cond-Belldiag}
\end{eqnarray}
We can use the above condition if the state $\rho_{\mathrm{Bell}}$ is separable,
that is for $a_+,a_-,b_+,b_-\leq {1\over 2}$. We have searched for 
the gap via $5\times 10^5$ random  tests  of $\rho_{\mathrm{Bell}}$ states, yet did not find any case with a gap in Eq. (\ref{eq:gap}). Indeed, for a large region of parameters we
are able to confirm, that considered states escape our technique. 

To see this, let us denote:
$a_+ = {\alpha_1 \over 2}, a_-={\alpha_2\over 2}, b_+={\alpha_3\over 2}, b_-={\alpha_4\over 2}$. Then
the condition of Eq. (\ref{eq:cond-Belldiag}) reads:
\begin{equation}
    2H\left(\left\{{\alpha_i\over 2}\right\}_{i=1}^4\right) - 2 > H\left(\left\{{(1-\alpha_i)\over 2}\right\}_{i=1}^4\right) 
\end{equation}
It turns out that the {\it converse}
inequality holds, if only $\alpha_i \notin [1/3,1/2]$ for all $i=1,\ldots,4$. This can be seen
from expanding $2 = \sum_i \alpha_i$,
and observing that the converse inequality holds element-wise:
\begin{equation}
    2 \eta\left({\alpha_i\over 2}\right) - \alpha_i\leq \eta\left({{1-\alpha_i}\over 2 }\right),
\end{equation}
under considered condition on $\alpha_i$, where $\eta(x)=-x\log_2 x$. The latter fact is confirmed by plotting the difference of l.h.s. and r.h.s. using Mathematica 7.0. In terms of parameters $a_\pm$ and $b_\pm$ of the state $\rho_{\mathrm{BellG}}$ we can not decide based on aforementioned results if the state has limited repeated randomness if $a_\pm,b_\pm \in [0,{1\over 6})\cup ({1\over 4},{1\over 2}]$. This fact allows us to conjecture, that all the states $\rho_{\mathrm{BellG}}$ escape our technique. 

\section{Discussion}
\label{sec:discussion}
In this manuscript we have studied relationship between private key  and private randomness obtainable from quantum states, treated as quantum resources. We have shown that the states containing ideal privacy (private dits) belong to the set of states containing ideal private randomness (independent dits). We have then asked if the topology of loyalty in network of repeaters can be modified by free operations of the resource theory of private randomness. We focused on the simplest repeater: two stations $A$ and $B$ linked by connections with an intermediate station $C$. The problem we focused on is whether there exists such an action of the three parties that, after performing it, $A$ can relay solely on loyalty of $B$ instead of trusting an intermediate party $C$. While entanglement swapping is an example of such type of an action in the case of  pure (maximally entangled) states, we show that in case of the mixed states it is not so (in general).

To achieve our goal, in analogy to the rate of repeated private key, we have defined the rate of repeated private randomness and showed an upper bound on the latter quantity. It is equal to twice the relative entropy with respect to the maximally mixed state. The bound holds for states with positive partial transposition. To exemplify the phenomenon, we showed that the separable Werner states for sufficiently large dimensions exhibit a gap between localisable private randomness and repeated one. Interestingly, the states used in \cite{BCHW}, exhibiting limitation on the repeated key, can not serve as good examples in our context. This is due to the factor $2$ appearing in our upper bound (one cannot achieve the gap between $I(A:B)_\rho$ and $2I(A:B)_{\rho^\Gamma}$). Improving the bound to characterize the subset of states (especially the subset of separable ones) that exhibit the gap between private and repeated private randomness is an important direction to study.  Our Theorem \ref{thm:simple-bound} and Corollary \ref{cor:simple-bound}, are analogues of Lemma $12$ and Theorem $13$ of \cite{BCHW}, respectively. The former yield the same bound as the one presented in our main result (Corollary \ref{cor:boundsRelEnt}). This is in contrast with the results for the private key.  Indeed, in the latter case, the mentioned Theorem $13$ of \cite{BCHW}, presents the bound on repeated private key {\it without} factor $2$.  However, a study in this direction allowed us to show that for PPT states the repeated private randomness of $\rho\otimes{\tilde \rho}$ is upper bounded by the same function evaluated on $\rho^{\Gamma} \otimes {\tilde \rho}^{\Gamma}$, which is of independent interest.

We also studied a limited repeater of private randomness in which the three parties first perform identical operations on each copy, and later perform the best $\mathrm{CLODCC}$ protocol on all obtained copies of $A$ and $B$, without the help of $C$. We showed that a certain idit, which is not in PPT set, exhibits an extreme gap for large $d$. Our findings in this respect do not have a direct analogue in \cite{BCHW}, and can be extended to hold for a private key.

Presented results open an interesting perspective for further research. 
First of all one could discuss the implication of results presented in the paper for the simplest possible case, i.e. $2\otimes 2$ states. The first step toward solution has been made in Section~\ref{2qubitsexamples},  showing that such construction is not straightforward and more sophisticated techniques or candidates are needed.

Secondly, as it was proposed also in \cite{YHW}, one could consider in the context of our paper {\it the amortised} approach in which the allowed operations can bring $k$ bits of private randomness (e.g. in form of purity). The output randomness gets further lowered by $k$ in the end. This is to compute the private randomness content of a given quantum state rather then private randomness of an operation. Since the latter class of operation is still to be explored, we have followed here the approach of \cite{YHW} based on CLODCC operations.

From the broader perspective  we could ask a question: {\it which quantum resources (or just properties of quantum states) are ``transferable'' via quantum
network of mixed states}? We have shown that the limitation on the transfer of certain resources is not bound to private key only. Designing axioms for a resource theory to have limited transfer is an interesting direction of studies.

 It is also essential to show an analog of the  obtained results for channels rather than states, in the spirit of \cite{Christandl-Hermes}, and for states with negative partial transposition, adopting methods of \cite{FerraraChristandl}. Further investigation of inter-dependencies between private randomness and private key can also lead to fruitful results.

\section*{Acknowledgements}
{KH thanks Pawe\l{} Horodecki for discussion on possible axiomatic approach to presented results, Dong Yang for proof reading of an early draft of this manuscript and Andreas Winter for sharing an observation that singlet is an ibit. KH, RPK and RS acknowledge support of the National Science Centre grant Sonata Bis 5 (grant no. 2015/18/E/ST2/00327) from the National Science Center. KH acknowledge partial support by the Foundation for Polish Science (IRAP project, ICTQT, contract no. 2018/MAB/5, co-financed by EU via Smart Growth Operational Programme).

\bibliographystyle{apsrev}
\bibliography{References-2a}

\begin{thebibliography}{31}
\expandafter\ifx\csname natexlab\endcsname\relax\def\natexlab#1{#1}\fi
\expandafter\ifx\csname bibnamefont\endcsname\relax
  \def\bibnamefont#1{#1}\fi
\expandafter\ifx\csname bibfnamefont\endcsname\relax
  \def\bibfnamefont#1{#1}\fi
\expandafter\ifx\csname citenamefont\endcsname\relax
  \def\citenamefont#1{#1}\fi
\expandafter\ifx\csname url\endcsname\relax
  \def\url#1{\texttt{#1}}\fi
\expandafter\ifx\csname urlprefix\endcsname\relax\def\urlprefix{URL }\fi
\providecommand{\bibinfo}[2]{#2}
\providecommand{\eprint}[2][]{\url{#2}}

\bibitem[{\citenamefont{Kozlowski and Wehner}(2019)}]{Kozlowski-Wehner}
\bibinfo{author}{\bibfnamefont{W.}~\bibnamefont{Kozlowski}} \bibnamefont{and}
  \bibinfo{author}{\bibfnamefont{S.}~\bibnamefont{Wehner}}, in
  \emph{\bibinfo{booktitle}{Proceedings of the Sixth Annual ACM International
  Conference on Nanoscale Computing and Communication - {NANOCOM} 19}}
  (\bibinfo{publisher}{ACM Press}, \bibinfo{year}{2019}),
  \urlprefix\url{https://doi.org/10.1145/3345312.3345497}.

\bibitem[{\citenamefont{\ifmmode~\dot{Z}\else \.{Z}\fi{}ukowski
  et~al.}(1993)\citenamefont{\ifmmode~\dot{Z}\else \.{Z}\fi{}ukowski,
  Zeilinger, Horne, and Ekert}}]{ent_swapping}
\bibinfo{author}{\bibfnamefont{M.}~\bibnamefont{\ifmmode~\dot{Z}\else
  \.{Z}\fi{}ukowski}},
  \bibinfo{author}{\bibfnamefont{A.}~\bibnamefont{Zeilinger}},
  \bibinfo{author}{\bibfnamefont{M.}~\bibnamefont{Horne}}, \bibnamefont{and}
  \bibinfo{author}{\bibfnamefont{A.}~\bibnamefont{Ekert}},
  \bibinfo{journal}{Phys. Rev. Lett.} \textbf{\bibinfo{volume}{71}},
  \bibinfo{pages}{4287} (\bibinfo{year}{1993}),
  \urlprefix\url{https://link.aps.org/doi/10.1103/PhysRevLett.71.4287}.

\bibitem[{\citenamefont{Bennett et~al.}(1996)\citenamefont{Bennett, Brassard,
  Popescu, Schumacher, Smolin, and Wootters}}]{BBPSSW1996}
\bibinfo{author}{\bibfnamefont{C.}~\bibnamefont{Bennett}},
  \bibinfo{author}{\bibfnamefont{G.}~\bibnamefont{Brassard}},
  \bibinfo{author}{\bibfnamefont{S.}~\bibnamefont{Popescu}},
  \bibinfo{author}{\bibfnamefont{B.}~\bibnamefont{Schumacher}},
  \bibinfo{author}{\bibfnamefont{J.~A.} \bibnamefont{Smolin}},
  \bibnamefont{and} \bibinfo{author}{\bibfnamefont{W.~K.}
  \bibnamefont{Wootters}}, \bibinfo{journal}{Phys. Rev. Lett.}
  \textbf{\bibinfo{volume}{76}}, \bibinfo{pages}{722} (\bibinfo{year}{1996}),
  \eprint{quant-ph/9511027}.

\bibitem[{\citenamefont{Briegel et~al.}(1998)\citenamefont{Briegel, D\"ur,
  Cirac, and Zoller}}]{repeatersPRL}
\bibinfo{author}{\bibfnamefont{H.-J.} \bibnamefont{Briegel}},
  \bibinfo{author}{\bibfnamefont{W.}~\bibnamefont{D\"ur}},
  \bibinfo{author}{\bibfnamefont{J.}~\bibnamefont{Cirac}}, \bibnamefont{and}
  \bibinfo{author}{\bibfnamefont{P.}~\bibnamefont{Zoller}},
  \bibinfo{journal}{Phys. Rev. Lett.} \textbf{\bibinfo{volume}{81}},
  \bibinfo{pages}{5932} (\bibinfo{year}{1998}),
  \urlprefix\url{https://link.aps.org/doi/10.1103/PhysRevLett.81.5932}.

\bibitem[{\citenamefont{D\"ur et~al.}(1999)\citenamefont{D\"ur, Briegel, Cirac,
  and Zoller}}]{repeaters}
\bibinfo{author}{\bibfnamefont{W.}~\bibnamefont{D\"ur}},
  \bibinfo{author}{\bibfnamefont{H.-J.} \bibnamefont{Briegel}},
  \bibinfo{author}{\bibfnamefont{J.}~\bibnamefont{Cirac}}, \bibnamefont{and}
  \bibinfo{author}{\bibfnamefont{P.}~\bibnamefont{Zoller}},
  \bibinfo{journal}{Phys. Rev. A} \textbf{\bibinfo{volume}{59}},
  \bibinfo{pages}{169} (\bibinfo{year}{1999}),
  \urlprefix\url{https://link.aps.org/doi/10.1103/PhysRevA.59.169}.

\bibitem[{\citenamefont{Horodecki
  et~al.}(2009{\natexlab{a}})\citenamefont{Horodecki, Horodecki, Horodecki, and
  Horodecki}}]{RMPK-quant-ent}
\bibinfo{author}{\bibfnamefont{R.}~\bibnamefont{Horodecki}},
  \bibinfo{author}{\bibfnamefont{P.}~\bibnamefont{Horodecki}},
  \bibinfo{author}{\bibfnamefont{M.}~\bibnamefont{Horodecki}},
  \bibnamefont{and}
  \bibinfo{author}{\bibfnamefont{K.}~\bibnamefont{Horodecki}},
  \bibinfo{journal}{Reviews of Modern Physics} \textbf{\bibinfo{volume}{81}},
  \bibinfo{pages}{865} (\bibinfo{year}{2009}{\natexlab{a}}), ISSN
  \bibinfo{issn}{1539-0756}, \eprint{quant-ph/07022205v2},
  \urlprefix\url{http://dx.doi.org/10.1103/RevModPhys.81.865}.

\bibitem[{\citenamefont{{B{\"a}uml} et~al.}(2015)\citenamefont{{B{\"a}uml},
  {Christandl}, {Horodecki}, and {Winter}}}]{BCHW}
\bibinfo{author}{\bibfnamefont{S.}~\bibnamefont{{B{\"a}uml}}},
  \bibinfo{author}{\bibfnamefont{M.}~\bibnamefont{{Christandl}}},
  \bibinfo{author}{\bibfnamefont{K.}~\bibnamefont{{Horodecki}}},
  \bibnamefont{and} \bibinfo{author}{\bibfnamefont{A.}~\bibnamefont{{Winter}}},
  \bibinfo{journal}{Nature Communications} \textbf{\bibinfo{volume}{6}},
  \bibinfo{eid}{6908} (\bibinfo{year}{2015}), \eprint{1402.5927}.

\bibitem[{\citenamefont{Horodecki
  et~al.}(2005{\natexlab{a}})\citenamefont{Horodecki, Horodecki, Horodecki, and
  Oppenheim}}]{pptkey}
\bibinfo{author}{\bibfnamefont{K.}~\bibnamefont{Horodecki}},
  \bibinfo{author}{\bibfnamefont{M.}~\bibnamefont{Horodecki}},
  \bibinfo{author}{\bibfnamefont{P.}~\bibnamefont{Horodecki}},
  \bibnamefont{and}
  \bibinfo{author}{\bibfnamefont{J.}~\bibnamefont{Oppenheim}},
  \bibinfo{journal}{Phys. Rev. Lett.} \textbf{\bibinfo{volume}{94}},
  \bibinfo{pages}{160502} (\bibinfo{year}{2005}{\natexlab{a}}),
  \urlprefix\url{https://link.aps.org/doi/10.1103/PhysRevLett.94.160502}.

\bibitem[{\citenamefont{Horodecki
  et~al.}(2009{\natexlab{b}})\citenamefont{Horodecki, Horodecki, Horodecki, and
  Oppenheim}}]{keyhuge}
\bibinfo{author}{\bibfnamefont{K.}~\bibnamefont{Horodecki}},
  \bibinfo{author}{\bibfnamefont{M.}~\bibnamefont{Horodecki}},
  \bibinfo{author}{\bibfnamefont{P.}~\bibnamefont{Horodecki}},
  \bibnamefont{and}
  \bibinfo{author}{\bibfnamefont{J.}~\bibnamefont{Oppenheim}},
  \bibinfo{journal}{IEEE Transactions on Information Theory}
  \textbf{\bibinfo{volume}{55}}, \bibinfo{pages}{1898}
  (\bibinfo{year}{2009}{\natexlab{b}}), ISSN \bibinfo{issn}{1557-9654},
  \urlprefix\url{http://dx.doi.org/10.1109/TIT.2008.2009798}.

\bibitem[{\citenamefont{Bera et~al.}(2017)\citenamefont{Bera, Ac{\'{\i}}n,
  Ku{\'{s}}, Mitchell, and Lewenstein}}]{Randomness-review}
\bibinfo{author}{\bibfnamefont{M.}~\bibnamefont{Bera}},
  \bibinfo{author}{\bibfnamefont{A.}~\bibnamefont{Ac{\'{\i}}n}},
  \bibinfo{author}{\bibfnamefont{M.}~\bibnamefont{Ku{\'{s}}}},
  \bibinfo{author}{\bibfnamefont{M.}~\bibnamefont{Mitchell}}, \bibnamefont{and}
  \bibinfo{author}{\bibfnamefont{M.}~\bibnamefont{Lewenstein}},
  \bibinfo{journal}{Reports on Progress in Physics}
  \textbf{\bibinfo{volume}{80}}, \bibinfo{pages}{124001}
  (\bibinfo{year}{2017}),
  \urlprefix\url{https://doi.org/10.1088/1361-6633/aa8731}.

\bibitem[{\citenamefont{Berta et~al.}(2014)\citenamefont{Berta, Fawzi, and
  Wehner}}]{BFW-decoupling}
\bibinfo{author}{\bibfnamefont{M.}~\bibnamefont{Berta}},
  \bibinfo{author}{\bibfnamefont{O.}~\bibnamefont{Fawzi}}, \bibnamefont{and}
  \bibinfo{author}{\bibfnamefont{S.}~\bibnamefont{Wehner}},
  \bibinfo{journal}{IEEE} \textbf{\bibinfo{volume}{60}}, \bibinfo{pages}{1168}
  (\bibinfo{year}{2014}), \eprint{arXiv:1111.2026}.

\bibitem[{idq()}]{idq}
\bibinfo{note}{{h}ttps://www.idquantique.com/}.

\bibitem[{\citenamefont{D.Yang et~al.}(2019)\citenamefont{D.Yang, Horodecki,
  and Winter}}]{YHW}
\bibinfo{author}{\bibnamefont{D.Yang}},
  \bibinfo{author}{\bibfnamefont{K.}~\bibnamefont{Horodecki}},
  \bibnamefont{and} \bibinfo{author}{\bibfnamefont{A.}~\bibnamefont{Winter}},
  \bibinfo{journal}{Physical Review Letters} \textbf{\bibinfo{volume}{123}}
  (\bibinfo{year}{2019}),
  \urlprefix\url{https://doi.org/10.1103/physrevlett.123.170501}.

\bibitem[{\citenamefont{{Chitambar} and {Gour}}(2019)}]{Chitambar-Gour}
\bibinfo{author}{\bibfnamefont{E.}~\bibnamefont{{Chitambar}}} \bibnamefont{and}
  \bibinfo{author}{\bibfnamefont{G.}~\bibnamefont{{Gour}}},
  \bibinfo{journal}{Reviews of Modern Physics} \textbf{\bibinfo{volume}{91}},
  \bibinfo{eid}{025001} (\bibinfo{year}{2019}).

\bibitem[{\citenamefont{Peres}(1993)}]{Peres-book}
\bibinfo{author}{\bibfnamefont{A.}~\bibnamefont{Peres}},
  \emph{\bibinfo{title}{{"Quantum Theory: Concepts and Methods"}}}
  (\bibinfo{publisher}{Kluwer, Dordrecht}, \bibinfo{year}{1993}).

\bibitem[{\citenamefont{Horodecki
  et~al.}(2005{\natexlab{b}})\citenamefont{Horodecki, Horodecki, Horodecki,
  Oppenheim, {Sen(De)}, Sen, and Synak-Radtke}}]{huge-delta}
\bibinfo{author}{\bibfnamefont{M.}~\bibnamefont{Horodecki}},
  \bibinfo{author}{\bibfnamefont{P.}~\bibnamefont{Horodecki}},
  \bibinfo{author}{\bibfnamefont{R.}~\bibnamefont{Horodecki}},
  \bibinfo{author}{\bibfnamefont{J.}~\bibnamefont{Oppenheim}},
  \bibinfo{author}{\bibfnamefont{A.}~\bibnamefont{{Sen(De)}}},
  \bibinfo{author}{\bibfnamefont{U.}~\bibnamefont{Sen}}, \bibnamefont{and}
  \bibinfo{author}{\bibfnamefont{B.}~\bibnamefont{Synak-Radtke}},
  \bibinfo{journal}{Phys. Rev. A} \textbf{\bibinfo{volume}{71}},
  \bibinfo{pages}{062307} (\bibinfo{year}{2005}{\natexlab{b}}),
  \eprint{quant-ph/0410090}.

\bibitem[{\citenamefont{Horodecki et~al.}(2008)\citenamefont{Horodecki,
  Pankowski, Horodecki, and Horodecki}}]{smallkey}
\bibinfo{author}{\bibfnamefont{K.}~\bibnamefont{Horodecki}},
  \bibinfo{author}{\bibfnamefont{{\L}.}~\bibnamefont{Pankowski}},
  \bibinfo{author}{\bibfnamefont{M.}~\bibnamefont{Horodecki}},
  \bibnamefont{and}
  \bibinfo{author}{\bibfnamefont{P.}~\bibnamefont{Horodecki}},
  \bibinfo{journal}{IEEE Transactions on Information Theory}
  \textbf{\bibinfo{volume}{54}}, \bibinfo{pages}{2621} (\bibinfo{year}{2008}),
  ISSN \bibinfo{issn}{1557-9654},
  \urlprefix\url{http://dx.doi.org/10.1109/TIT.2008.921709}.

\bibitem[{\citenamefont{Werner}(1989)}]{Werner1989}
\bibinfo{author}{\bibfnamefont{R.}~\bibnamefont{Werner}},
  \bibinfo{journal}{Phys. Rev. A} \textbf{\bibinfo{volume}{40}},
  \bibinfo{pages}{4277} (\bibinfo{year}{1989}).

\bibitem[{\citenamefont{Oppenheim et~al.}(2002)\citenamefont{Oppenheim,
  Horodecki, Horodecki, and Horodecki}}]{OHHH2001}
\bibinfo{author}{\bibfnamefont{J.}~\bibnamefont{Oppenheim}},
  \bibinfo{author}{\bibfnamefont{M.}~\bibnamefont{Horodecki}},
  \bibinfo{author}{\bibfnamefont{P.}~\bibnamefont{Horodecki}},
  \bibnamefont{and}
  \bibinfo{author}{\bibfnamefont{R.}~\bibnamefont{Horodecki}},
  \bibinfo{journal}{Physical Review Letters} \textbf{\bibinfo{volume}{89}}
  (\bibinfo{year}{2002}), ISSN \bibinfo{issn}{1079-7114},
  \urlprefix\url{http://dx.doi.org/10.1103/PhysRevLett.89.180402}.

\bibitem[{\citenamefont{{Streltsov} et~al.}(2017)\citenamefont{{Streltsov},
  {Adesso}, and {Plenio}}}]{Alex-Review}
\bibinfo{author}{\bibfnamefont{A.}~\bibnamefont{{Streltsov}}},
  \bibinfo{author}{\bibfnamefont{G.}~\bibnamefont{{Adesso}}}, \bibnamefont{and}
  \bibinfo{author}{\bibfnamefont{M.}~\bibnamefont{{Plenio}}},
  \bibinfo{journal}{Reviews of Modern Physics} \textbf{\bibinfo{volume}{89}},
  \bibinfo{eid}{041003} (\bibinfo{year}{2017}), \eprint{1609.02439}.

\bibitem[{\citenamefont{Curty and L\"{u}tkenhaus}(2004)}]{CurtyLewLut}
\bibinfo{author}{\bibfnamefont{M.}~\bibnamefont{Curty},
  \bibfnamefont{M.~Lewenstein}} \bibnamefont{and}
  \bibinfo{author}{\bibfnamefont{N.}~\bibnamefont{L\"{u}tkenhaus}},
  \bibinfo{journal}{Physical Review Letters} \textbf{\bibinfo{volume}{92}}
  (\bibinfo{year}{2004}), ISSN \bibinfo{issn}{1079-7114},
  \urlprefix\url{http://dx.doi.org/10.1103/PhysRevLett.92.217903}.

\bibitem[{\citenamefont{Horodecki et~al.}(2006)\citenamefont{Horodecki, Leung,
  Lo, and Oppenheim}}]{bigkey}
\bibinfo{author}{\bibfnamefont{K.}~\bibnamefont{Horodecki}},
  \bibinfo{author}{\bibfnamefont{D.}~\bibnamefont{Leung}},
  \bibinfo{author}{\bibfnamefont{H.}~\bibnamefont{Lo}}, \bibnamefont{and}
  \bibinfo{author}{\bibfnamefont{J.}~\bibnamefont{Oppenheim}},
  \bibinfo{journal}{Physical Review Letters} \textbf{\bibinfo{volume}{96}}
  (\bibinfo{year}{2006}), ISSN \bibinfo{issn}{1079-7114},
  \urlprefix\url{http://dx.doi.org/10.1103/PhysRevLett.96.070501}.

\bibitem[{\citenamefont{Piani}(2009)}]{Piani2009-relent}
\bibinfo{author}{\bibfnamefont{M.}~\bibnamefont{Piani}},
  \bibinfo{journal}{Physical Review Letters} \textbf{\bibinfo{volume}{103}}
  (\bibinfo{year}{2009}), ISSN \bibinfo{issn}{1079-7114},
  \urlprefix\url{http://dx.doi.org/10.1103/PhysRevLett.103.160504}.

\bibitem[{\citenamefont{Alicki and Fannes}(2004)}]{Alicki_2004}
\bibinfo{author}{\bibfnamefont{R.}~\bibnamefont{Alicki}} \bibnamefont{and}
  \bibinfo{author}{\bibfnamefont{M.}~\bibnamefont{Fannes}},
  \bibinfo{journal}{Journal of Physics A: Mathematical and General}
  \textbf{\bibinfo{volume}{37}}, \bibinfo{pages}{L55} (\bibinfo{year}{2004}),
  \urlprefix\url{https://doi.org/10.1088%2F0305-4470%2F37%2F5%2Fl01}.

\bibitem[{\citenamefont{Shirokov}(2017)}]{Shirokov_2017}
\bibinfo{author}{\bibfnamefont{M.}~\bibnamefont{Shirokov}},
  \bibinfo{journal}{Journal of Mathematical Physics}
  \textbf{\bibinfo{volume}{58}}, \bibinfo{pages}{102202}
  (\bibinfo{year}{2017}), ISSN \bibinfo{issn}{1089-7658},
  \urlprefix\url{http://dx.doi.org/10.1063/1.4987135}.

\bibitem[{\citenamefont{Schumacher}(1995)}]{Schumacher1995}
\bibinfo{author}{\bibfnamefont{B.}~\bibnamefont{Schumacher}},
  \bibinfo{journal}{Phys. Rev. A} \textbf{\bibinfo{volume}{51}},
  \bibinfo{pages}{2738} (\bibinfo{year}{1995}),
  \urlprefix\url{https://link.aps.org/doi/10.1103/PhysRevA.51.2738}.

\bibitem[{\citenamefont{Nielsen and Chuang}(2000)}]{Nielsen-Chuang}
\bibinfo{author}{\bibfnamefont{M.}~\bibnamefont{Nielsen}} \bibnamefont{and}
  \bibinfo{author}{\bibfnamefont{I.}~\bibnamefont{Chuang}},
  \emph{\bibinfo{title}{{"Quantum Computation and Quantum Information"}}}
  (\bibinfo{publisher}{Cambridge University Press,Cambridge},
  \bibinfo{year}{2000}).

\bibitem[{\citenamefont{Werner}(2001)}]{werner-alltel}
\bibinfo{author}{\bibfnamefont{R.}~\bibnamefont{Werner}}, \bibinfo{journal}{J.
  Phys. A: Math. Gen.} \textbf{\bibinfo{volume}{34}}, \bibinfo{pages}{7081}
  (\bibinfo{year}{2001}), \eprint{quant-ph/0003070}.

\bibitem[{\citenamefont{Bengtsson and
  \.Zyczkowski}(2006)}]{BengtssonZyczkowski-book}
\bibinfo{author}{\bibfnamefont{I.}~\bibnamefont{Bengtsson}} \bibnamefont{and}
  \bibinfo{author}{\bibfnamefont{K.}~\bibnamefont{\.Zyczkowski}},
  \emph{\bibinfo{title}{{{Geometry of Quantum States. An Introduction to
  Quantum Entanglement}}}} (\bibinfo{publisher}{Cambridge University Press},
  \bibinfo{year}{2006}).

\bibitem[{\citenamefont{{Christandl} and
  {M{\"u}ller-Hermes}}(2017)}]{Christandl-Hermes}
\bibinfo{author}{\bibfnamefont{M.}~\bibnamefont{{Christandl}}}
  \bibnamefont{and}
  \bibinfo{author}{\bibfnamefont{A.}~\bibnamefont{{M{\"u}ller-Hermes}}},
  \bibinfo{journal}{Communications in Mathematical Physics}
  \textbf{\bibinfo{volume}{353}}, \bibinfo{pages}{821} (\bibinfo{year}{2017}),
  \eprint{1604.03448}.

\bibitem[{\citenamefont{Christandl and Ferrara}(2017)}]{FerraraChristandl}
\bibinfo{author}{\bibfnamefont{M.}~\bibnamefont{Christandl}} \bibnamefont{and}
  \bibinfo{author}{\bibfnamefont{R.}~\bibnamefont{Ferrara}},
  \bibinfo{journal}{Physical Review Letters} \textbf{\bibinfo{volume}{119}}
  (\bibinfo{year}{2017}),
  \urlprefix\url{https://doi.org/10.1103/physrevlett.119.220506}.

\end{thebibliography}
\end{document}